\documentclass[sigplan,screen]{acmart}

\usepackage[]{hyperref}

\usepackage{tikz}
\usepackage{amsmath}

\usepackage{filecontents}
\usepackage[normalem]{ulem}
\usepackage{graphicx}

\usepackage{subcaption}
\usepackage{flushend}

\usepackage{multirow}
\usepackage{rotating}
\usepackage{array}
\usepackage{balance}
\usepackage{enumitem}
\usepackage[ruled,linesnumbered,nofillcomment]{algorithm2e}
\usepackage{balance}

\usepackage[subtle]{savetrees}
\usepackage{verbatim}
\usepackage{soul}
\usepackage{listings}
\usepackage{caption}
\usepackage{xcolor}
\usepackage{afterpage}
\newlength{\oldtextfloatsep}
\definecolor{darkcerulean}{rgb}{0.03, 0.27, 0.49}
\definecolor{codegreen}{rgb}{0,0.6,0}
\definecolor{codegray}{rgb}{0.5,0.5,0.5}
\definecolor{codepurple}{rgb}{0.58,0,0.82}
\definecolor{backcolour}{rgb}{0.95,0.95,0.92}
\definecolor{ballblue}{rgb}{0.13, 0.67, 0.8}
\definecolor{blue}{rgb}{0.0, 0.23, 0.84}
\definecolor{cobalt}{rgb}{0.0, 0.28, 0.67}
\definecolor{coolblack}{rgb}{0.0, 0.18, 0.39}
\definecolor{darkcerulean}{rgb}{0.03, 0.27, 0.49}
\definecolor{readgreen}{rgb}{0, 0.398, 0.199}
\definecolor{writepurple}{rgb}{0.906, 0.5195, 0.906}
\definecolor{compactpurple}{rgb}{0.48, 0, 0.5078}
\definecolor{lightgreen}{rgb}{0.832, 0.90625, 0.828125}
\definecolor{darkblue}{rgb}{0, 0.3125, 0.93359}
\definecolor{darkorange}{rgb}{0.8867, 0.34375, 0.0703125}
\definecolor{asparagus}{rgb}{0.53, 0.66, 0.42}
\definecolor{atomictangerine}{rgb}{1.0, 0.6, 0.4}
\definecolor{beaver}{rgb}{0.62, 0.51, 0.44}
\definecolor{camel}{rgb}{0.86, 0.7, 0.52}
\definecolor{flax}{rgb}{0.93, 0.86, 0.51}
\definecolor{lavenderpurple}{rgb}{0.59, 0.48, 0.71}
\definecolor{lightpastelpurple}{rgb}{0.69, 0.61, 0.85}
\definecolor{ballbluelight}{rgb}{0.33, 0.77, 0.95}
\definecolor{lightgreend}{rgb}{0.8, 0.87, 0.77}
\definecolor{writepurpled}{rgb}{0.92, 0.72, 0.92}

\lstdefinestyle{BashStyle} {
  frame=tb,
  language=bash,
  aboveskip=3mm,
  belowskip=3mm,
  showstringspaces=false,
  columns=flexible,
  basicstyle={\scriptsize\ttfamily},
  numbers=left,
  numbersep=5pt,
  numberstyle=\scriptsize\color{black},
  keywordstyle=\color{codepurple},
  commentstyle=\color{codegreen},
  stringstyle=\color{magenta},
  breaklines=true,
  breakatwhitespace=true,
  tabsize=2,
  classoffset=0,
  morekeywords={apt, pip3, make, cd, wget, python3, vim, update-grub, lsblk, git},
  keywordstyle=\color{codepurple},
  classoffset=1,
  morekeywords={sudo},
  keywordstyle=\color{blue},
  escapechar=`,
}

\usepackage{tikz}
\let\oldnl\nl
\newcommand{\nonl}{\renewcommand{\nl}{\let\nl\oldnl}}

\newcommand*\circlemr[1]{\tikz[baseline=(char.base)]{
            \node[shape=circle,draw=readgreen,inner sep=0.5pt,fill=readgreen,text=white, scale=0.7] (char) {#1};}}

\newcommand*\circlemw[1]{\tikz[baseline=(char.base)]{
            \node[shape=circle,draw=codepurple,inner sep=0.5pt ,fill=codepurple,text=white, scale=0.6] (char) {#1};}}
            
\newcommand*\circlems[1]{\tikz[baseline=(char.base)]{
            \node[shape=circle,draw=blue,inner sep=0.5pt,fill=blue,text=white, scale=0.7] (char) {#1};}}

\newcommand*\circleblank[1]{\tikz[baseline=(char.base)]{
            \node[shape=circle,draw=black,inner sep=0.5pt,fill=white,text=black, scale=0.85] (char) {#1};}}

\newcommand{\pname}[1]{{ByteFS}{#1}}

\DeclareRobustCommand{\hlcommon}[1]{#1}
\DeclareRobustCommand{\hlA}[1]{#1}
\DeclareRobustCommand{\hlB}[1]{#1}
\DeclareRobustCommand{\hlC}[1]{#1}
\DeclareRobustCommand{\hlD}[1]{#1}
\DeclareRobustCommand{\hlE}[1]{#1}

\copyrightyear{2025}
\acmYear{2025}
\setcopyright{rightsretained}
\acmConference[ASPLOS '25]{Proceedings of the 30th ACM International Conference on Architectural Support for Programming Languages and Operating Systems, Volume 1}{March 30--April 3, 2025}{Rotterdam, Netherlands}
\acmBooktitle{Proceedings of the 30th ACM International Conference on Architectural Support for Programming Languages and Operating Systems, Volume 1 (ASPLOS '25), March 30--April 3, 2025, Rotterdam, Netherlands}
\acmDOI{10.1145/3669940.3707250}
\acmISBN{979-8-4007-0698-1/25/03}

\makeatletter
\gdef\@copyrightpermission{
  \begin{minipage}{0.3\columnwidth}
   \href{https://urldefense.com/v3/__https://creativecommons.org/licenses/by-nc-sa/4.0/}{\includegraphics[width=0.90\textwidth]{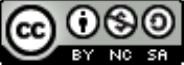}}
  \end{minipage}\hfill
  \begin{minipage}{0.7\columnwidth}
   \href{https://creativecommons.org/licenses/by-nc-sa/4.0/}{This work is licensed under a Creative Commons Attribution-NonCommercial-ShareAlike International 4.0 License.}
  \end{minipage}
  \vspace{5pt}
}
\makeatother



\begin{document}
\pagenumbering{gobble}

\title[\pname{}: System Support for (CXL-based) Memory-Semantic Solid-State Drives]{\pname{}: System Support for (CXL-based)\\ Memory-Semantic Solid-State Drives}

\author{Shaobo Li}
\authornote{Co-primary authors.}
\email{shaobol2@illinois.edu}
\affiliation{%
  \institution{University of Illinois Urbana-Champaign}
  \streetaddress{}
  \city{Urbana}
  \state{IL}
  \country{USA}
  \postcode{}
}

\author{Yirui Eric Zhou}
\authornotemark[1]
\email{yiruiz2@illinois.edu}
\affiliation{%
  \institution{University of Illinois Urbana-Champaign}
  \streetaddress{}
  \city{Urbana}
  \state{IL}
  \country{USA}
  \postcode{}
}

\author{Hao Ren}
\email{haor2@illinois.edu}
\affiliation{%
  \institution{University of Illinois Urbana-Champaign}
  \streetaddress{}
  \city{Urbana}
  \state{IL}
  \country{USA}
  \postcode{}
}

\author{Jian Huang}
\email{jianh@illinois.edu}
\affiliation{%
  \institution{University of Illinois Urbana-Champaign}
  \city{Urbana}
    \state{IL}
  \country{USA}
}

\begin{abstract}


Unlike non-volatile memory that resides on the processor memory bus, memory-semantic  
solid-state drives (SSDs) support both byte and block access granularity via PCIe or CXL
interconnects. 
They provide scalable memory capacity using NAND flash at a much lower cost. In addition, they have different performance characteristics for their dual byte/block interface respectively, while offering essential memory semantics for upper-level software.
Such a byte-accessible storage device provides new implications on the software system design.

In this paper, we develop a new file system, named \pname{}, by rethinking the design 
primitives of file systems and SSD firmware to exploit the advantages of both byte and block-granular data accesses. \pname{} supports byte-granular data persistence to retain the persistence nature of SSDs. It extends the core data structure of file systems by enabling dual byte/block-granular data accesses. To facilitate the 
support for byte-granular writes, \pname{} manages the internal DRAM of 
SSD firmware in a log-structured manner and enables data coalescing to reduce the unnecessary I/O traffic to flash chips. 
\pname{} also enables coordinated data caching between the host page cache and SSD cache for best utilizing the precious memory resource. 
We implement \pname{} on both a real programmable SSD and an emulated memory-semantic SSD for sensitivity study. Compared to state-of-the-art file systems for non-volatile memory and conventional SSDs, \pname{} outperforms them by up to 2.7$\times$, while preserving the 
essential properties of a file system. \pname{} also reduces the write traffic to SSDs by up to 5.1$\times$ by alleviating unnecessary writes caused by both metadata and data updates in file systems. 

\end{abstract}

\maketitle

\section{Introduction}
\label{sec:intro}

Modern computer systems have an increasing demand for storage-class memory (SCM)~\cite{bpfs,www:scm}, as it promises many advantages such as scalable storage capacity, byte-addressable data accesses, and data durability. A typical example is non-volatile memory (NVM), which attracted much attention over the past decade~\cite{Qureshi-isca09,Starta,nvm:sosp23, trio:sosp23}. However, due to the insufficient supply of mature products on the market~\cite{intel-nvm}, we have to seek alternative practical memory technologies. 

Thanks to the byte-accessibility of the PCIe interconnect (e.g., NVMe and CXL), the commodity flash-based SSDs can provide byte-addressable data accesses by utilizing the memory-mapped I/O interface (MMIO)~\cite{flatflash,2bssd,cxlssd}.  
Therefore, we can access SSDs with both byte and block granularity. We call them as \textit{memory-semantic SSDs} (M-SSD). Unlike NVM devices such as Intel Optane Persistent Memory that use PCM-like storage medium and reside on the processor memory bus~\cite{intelnvm:micro2020}, the M-SSD is PCIe attached and built upon the mature NAND flash technology, which offers unique advantages. 
First, it relies solely on the byte-accessibility support of the existing PCIe interface, thus a majority of the commodity PCIe-attached SSDs can be reformed as memory-semantic SSDs. For instance, Samsung publicized its CXL-based SSDs built on the commodity NVMe SSDs recently~\cite{cxlssd}. Second, its backend storage medium NAND flash has much lower cost, 
and its storage capacity can easily scale up to terabytes per PCIe slot~\cite{flatflash}. Third, it provides the memory interface for software systems while preserving compelling storage properties such as data durability and simple deployment.


However, for decades, systems software like file systems are used to the generic block interface to access SSDs, resulting in high I/O amplification~\cite{io:apsys2017, flatflash}. They lack the support for dual byte and block interface for M-SSDs, causing sub-optimal performance and increased complexity of data management. File systems for persistent memory such as PMFS~\cite{pmfs} and NOVA~\cite{nova} focused on the byte interface, however, they were designed for NVM devices whose characteristics are fundamentally different from M-SSDs. Thus, none of the current file systems is a natural fit for M-SSDs. The M-SSD provides a set of unique opportunities and challenges for systems software.

We present \pname{}, an efficient file system for M-SSDs with software/hardware co-design. \pname{} transparently supports dual byte/block interface for high-performance data accesses, while preserving the essential properties of file systems, including crash consistency and data recovery. 
\pname{} extends the SSD firmware by managing its internal DRAM cache in a log-structured manner for accommodating the byte-granular data accesses, and enabling data coalescing to reduce I/O traffic to flash chips. It also enables coordinated data caching between the host page cache and the internal DRAM of the SSD. 

To develop \pname{}, we first conduct a thorough study of generic Linux file systems to understand the appropriate interface (i.e., byte or block) needed for core filesystem data structures for their interaction with the M-SSD, and their impact on I/O amplification. 
As we expected, some data structures like inode prefer byte-granular data access, and some structures like page cache prefer dual byte/block-granular data access depending on the data access pattern at runtime (see Table~\ref{tab:fsops}).
Our study ($\S$\ref{sec:study}) provides guidelines on developing \pname{}. 

\pname{} enables byte-granular persistent writes to reduce I/O amplification for a majority of data structures. For the data structures that prefer dual byte and block interfaces, such as page cache, data block, and data journal, \pname{} employs different policies to decide the appropriate data access granularity. As for the page cache, \pname{} utilizes the copy-on-write mechanism to track the writes to a page, and learn their data locality. \pname{} uses byte-granular writes for hot cachelines, and 
block-granular writes for a page with lower data locality. As for the writes to data blocks and data journal, \pname{} decides the write granularity based on the amount of data required for the durable writes to the SSD. It employs block-granular reads for all data structures to exploit the locality in the host cache and simplify their management.

Although M-SSDs enable the byte interface via PCIe/CXL, 
the flash chips inside the SSD support only page-granular accesses due to physical limitations~\cite{2bssd, flatflash}, causing extra I/O amplification. To address this challenge, \pname{} extends the SSD firmware by managing the internal SSD DRAM in a log-structured manner at cacheline granularity, and enabling data coalescing with background log cleaning. This reduces unnecessary I/O traffic caused by the mismatch of data access granularity between SSD DRAM (byte-granular) and flash chips (page-granular). \pname{} develops a lightweight indexing mechanism using skip lists for fast log lookup, and implements coordinated data caching between the host and SSD DRAM. Instead of caching flash pages in both SSD DRAM and host page cache, \pname{} caches flash pages only in the host page cache for best utilizing the precious SSD DRAM for persistent writes. 

\pname{} preserves essential file system properties, including crash consistency and data recovery. 
\pname{} develops a low-overhead transactional mechanism for filesystem operations using the byte-granular persistent writes and firmware-level logging. 
With the write log in the SSD DRAM, \pname{} facilitates the enforcement of crash consistency and data recovery. 

We implement \pname{} based on the Ext4 file system, and develop a full system prototype on a programmable SSD FPGA board to validate its functions and efficiency. We extend the PCIe protocol on the FPGA board to support byte-granular persistent writes, and modify the SSD firmware for the write log. We also develop an M-SSD emulator for sensitivity analysis. Compared to the file systems developed for NVM, such as 
PMFS~\cite{pmfs} and NOVA~\cite{nova}, and for block devices, such as Ext4, \pname{} improves the performance by up to 2.7$\times$. \pname{} also shows its friendliness to SSDs by reducing the write traffic by up to 5.1$\times$. In summary, we make the following contributions.

\begin{itemize}[leftmargin=*]

\item We conduct a characterization study of data access interface preferred by core data structures of file systems, it provides guidelines for enabling system software support for M-SSD.

\vspace{0.5ex}
\item We develop a new file system \pname{} for memory-semantic SSDs by supporting adaptive byte and block-granular data accesses, it significantly reduces the I/O amplification. 
\vspace{0.5ex}
\item We extend the SSD firmware to manage the SSD DRAM in a log-structured manner and enable data coalescing to reduce the I/O traffic caused by the mismatch of access granularity between SSD DRAM and flash chips. 

\vspace{0.5ex}
\item We present a coordinated data caching mechanism between the host page cache and SSD DRAM for 
best utilizing the precious SSD DRAM for writes. 
\item We utilize the firmware-level log-structured memory to speed up file system transaction commits, and leverage existing battery-backed SSD DRAM to retain transaction logs for fast data recovery.

\vspace{0.5ex}
\item We develop \pname{} with both a real programmable SSD and a memory-semantic SSD emulator, and demonstrate its efficiency with various filesystem benchmarks.

\end{itemize}

\section{Background and Motivation}
\label{sec:background}
We first discuss the technical background of M-SSD. And then, we discuss why its system support is highly desirable.

\begin{table}[t]
\caption{Characteristics of different memory devices. M-SSD represents the memory-semantic SSD. \hlD{M-SSD performance is measured with the M-SSD prototype built with the OpenSSD FPGA board.} Both SSD and M-SSD use PCIe 3.0 x4. DRAM and NVM are measured with single DIMM. }
\footnotesize
\begin{tabular}{|l|l|l|l|c|}
\hline
Memory & \begin{tabular}[c]{@{}l@{}}R/W Latency\\ (cacheline)\end{tabular} & \begin{tabular}[c]{@{}l@{}}Seq R/W \\ BW (4KB)\end{tabular} & \$/GB & Persistency \\ \hline
DRAM~\cite{Starta,skhynix}   & 100 ns  & 31.8 GB/s  &  \enspace8.6 & No      \\ \hline
NVM~\cite{izraelevitz2019basic,optanepm}     & 300/90 ns & 6.6/2.3 GB/s &  \enspace3.6  & Yes        \\ \hline
SSD~\cite{970pro}      & N/A & 3.5/2.5 GB/s & \enspace0.22 & Yes        \\ \hline
M-SSD~\cite{flatflash,2bssd}      & 4.8/0.6 $\mu$s & 3.5/2.5 GB/s & \textasciitilde0.22   & Yes        \\ \hline
\end{tabular}
\label{tab:device}
\end{table}

\subsection{Memory-Semantic SSDs}
\label{subsec:bytessd}
Storage devices like SSDs usually use the block interface to interact with upper-level system software like file systems. Recently, industry and academia have been exploiting the byte-addressability of SSDs~\cite{cxlssd, jung2022hotstorage, flatflash}. 
Unlike non-volatile memory (NVM)~\cite{chen2016review,bez2004non,optane} technologies, memory-semantic SSDs (M-SSDs) provide the byte interface by leveraging the in-device DRAM and the PCIe memory-mapped interface~\cite{pci}. By registering the device buffer address with the \textit{base address register} (BAR), the BIOS and OS discover and map the device buffer into the host memory space, and the host can access the mapped region via load/store instructions~\cite{2bssd,flatflash}. The in-device DRAM is used as a data buffer to serve byte-granular requests. \hlB{Battery-backed DRAM is used to assist data persistency upon power loss.} The M-SSD still keeps the normal block interface with conventional NVMe I/O commands, allowing the host to operate on dual byte and block interfaces. 


We compare the performance and cost of M-SSDs with other types of memory devices in Table~\ref{tab:device}. We measure the latency and bandwidth with real devices (see the experimental setup in $\S$\ref{subsec:eva_setup}). M-SSDs provide a 4.8$\mu$s read latency when reading cachelines in the device DRAM and offer a 0.6$\mu$s write latency with PCIe posted writes. With the new Compute Express Link (CXL)~\cite{cxl}, the latency can be further reduced.
However, it is still slower than NVMs that can achieve near-DRAM performance. For data accesses not served by the SSD DRAM, $\mu$s-level flash page accesses will result in a high delay.

Although the performance of M-SSDs might be less competitive compared to NVM or DRAM, they offer great cost-effectiveness. We obtained the lowest prices of these devices from popular memory vendors in 2024. The NVM price is derived from the Intel Optane Persistent Memory. Since we cannot obtain any price of M-SSD on the current market, we estimate it would be similar to SSDs, as they are developed upon commodity SSDs and have similar hardware components including the flash controller and flash chips. The cost would be possibly higher with CXL components, but compared to DRAM or NVM, M-SSDs are still significantly cheaper with a cost of around \$0.22/GB, making it a cost-effective solution in building high-performance storage systems.



\subsection{System Support for Memory-semantic SSDs}
\label{subsec:why}
Although M-SSDs show great potential in offering high performance and large capacity at a low cost, they lack system software support. A straightforward approach is to directly employ traditional file systems. However, they were designed for the block I/O interface. They manage underlying storage devices with fixed-size blocks (e.g., 512 bytes or 4 KB), and cannot utilize the byte-granular access brought by the M-SSD. We study the limitations of the block interface in detail in \S\ref{sec:study}. 


Apart from the traditional block-based file systems, there have been many emerging NVM file systems~\cite{bpfs,pmfs,nova,Starta,splitfs}. 
These file systems are all developed for byte-addressable NVM devices, which commonly leverage memory load/store interface to exploit ultra-low latency accesses. Reducing software overhead has been a critical design goal for such systems, and many utilize Direct Access techniques to bypass the host page cache or the OS kernel. However, na\"ively applying NVM file systems to M-SSDs is inadequate. The performance characteristics of M-SSDs differ from NVMs (\S\ref{subsec:bytessd}), as the high PCIe latency and inevitable flash accesses are still the bottleneck. 

\hlE{Therefore, without changing the inherent design,} it is hard to fully exploit the performance benefits of M-SSDs with existing file systems, as they all lack the support for both byte and block interface, and the consideration of the unique performance characteristics. This motivates us to develop a new system \pname{} for M-SSDs.

\begin{table}[t]
  \footnotesize
  \centering
  \caption{I/O amplification of using block I/O interface in conventional file systems Ext4 and F2FS.} 
  \label{tab:fs_amp}

\centering
\begin{tabular}{|ll|c|c|c|c|c|}
\hline
\multicolumn{2}{|l|}{}                              & \multicolumn{1}{l|}{Varmail} & \multicolumn{1}{l|}{Fileserver} & \multicolumn{1}{l|}{Webproxy} & \multicolumn{1}{l|}{Webserver} & \multicolumn{1}{l|}{OLTP} \\ \hline
\multicolumn{1}{|l|}{\multirow{2}{*}{\rotatebox{90}{Ext4\;}}} & Write & 3.85$\times$                 & 6.21$\times$                    & 1.43$\times$                  & 1.66$\times$                   & 2.17$\times$              \\ \cline{2-7} 
\multicolumn{1}{|l|}{}                      & Read  & 1.21$\times$                 & 1.15$\times$                    & 1.25$\times$                  & 1.71$\times$                   & 1.52$\times$              \\ \hline
\multicolumn{1}{|l|}{\multirow{2}{*}{\rotatebox{90}{F2FS\;}}} & Write & 2.14$\times$                 & 1.92$\times$                    & 1.67$\times$                  & 1.06$\times$                   & 1.10$\times$              \\ \cline{2-7} 
\multicolumn{1}{|l|}{}                      & Read  & 1.67$\times$                 & 1.42$\times$                    & 1.35$\times$                  & 1.18$\times$                   & 1.13$\times$              \\ \hline
\end{tabular}

\vspace{-2.5ex}
\end{table}

\newcommand{\specialcell}[2][c]{%
  \begin{tabular}[#1]{@{}c@{}}#2\end{tabular}}

\newcommand{\specialcellleft}[2][c]{%
  \begin{tabular}[#1]{@{}l@{}}#2\end{tabular}}

\begin{table*}[t]
  \scriptsize
  \caption{The core components in the generic Linux file systems (Ext4 and F2FS) studied in this paper.} 
  \label{tab:fsops}
  \vspace{-2ex}
  \center
  \begin{tabular}{|p{45pt}<{\centering}|p{90pt}<{}|p{110pt}<{}|p{155pt}<{}|p{42pt}<{\centering}|}
  \hline
  \textbf{\specialcell{Data \\ Structure}} & \textbf{Description} & \textbf{Related File Operations} & \textbf{Disk Access Frequency} & \textbf{\specialcell{Preferred\\Interface}}  \\
    \hline
      \specialcell{Superblock} & \specialcellleft{The metadata that defines the \\file system.} & 
      \specialcellleft{File system mount and remount.} & 
      \specialcellleft{R: Low, refers to the general info of a file system.\\ W: Low, infrequent global metadata update.} & 
      \specialcell{R: Block \\ W: Block}\\
    \hline
      \specialcell{Block List} & \specialcellleft{Marks free and used blocks in\\ the file system.} & 
      \specialcellleft{File create, append, and truncate.} & 
      \specialcellleft{R: Depends, the structure can be cached in host DRAM. \\ W: High, when blocks are allocated/freed.} & 
      \specialcell{R: Block \\ W: Byte} \\
    \hline
      \specialcell{Inode List} & \specialcellleft{Marks free and used inodes in\\ the inode table.} & 
      \specialcellleft{File create and unlink.} & 
      \specialcellleft{R: Depends, the structure can be cached in host DRAM. \\ W: High, when files are created and deleted.} & 
      \specialcell{R: Block \\ W: Byte} \\
    \hline
      \specialcell{Inode} & \specialcellleft{Describes a file system object\\ such as a file or a directory.} & 
      \specialcellleft{File/directory create, file rename, \\file truncate, file link/unlink, and others.} & 
      \specialcellleft{R: High, most operations involve file/dir metadata.\\ W: High, most operations involve metadata updates.} & 
      \specialcell{R: Block \\ W: Byte} \\
    \hline
      \specialcell{Data Pointer} & \specialcellleft{Indexes the location of the file \\ data on the storage device.} & 
      \specialcellleft{File read and write.} & 
      \specialcellleft{R: High, when the file system reads data from any file. \\ W: High, when the file system appends or truncates file. } & 
      \specialcell{R: Block \\ W: Byte} \\
    \hline
      \specialcell{Directory Entry} & \specialcellleft{Holds child inode information\\ under the directory.} & 
      \specialcellleft{File/dir create, rename, and link/unlink.} & 
      \specialcellleft{R: Depends, mainly on the file lookup. \\ W: Depends, higher when having more dirs and files.} & 
      \specialcell{R: Block \\ W: Byte} \\
    \hline
      \specialcell{Page Cache} & \specialcellleft{Transparent cache that exploits\\ data locality of applications.} & 
      \specialcellleft{File create, read, and write.} & 
      \specialcellleft{R: High, for data-intensive workloads. \\ W: High, for data-intensive workloads.} & 
      \specialcell{R: Block \\ W: Block/Byte} \\
    \hline
      \specialcell{Data Block} & \specialcellleft{File data stored in the file system.} & 
      \specialcellleft{File create, read, and write.} & 
      \specialcellleft{R: High, for data-intensive workloads. \\ W: High, for data-intensive workloads.} & 
      \specialcell{R: Block/Byte  \\ W: Block/Byte} \\
    \hline
      \specialcell{Data Journal} & \specialcellleft{History of operations executed in\\ the file system.} & 
      \specialcellleft{Operation that modifies critical \\ file system data structures.} & 
      \specialcellleft{R: Low, mostly accessed during recovery. \\ W: High, most operations that require journaling.} & 
      \specialcell{R: Block \\ W: Block/Byte} \\
    \hline
  \end{tabular}
\end{table*}




\section{A Quantitative Study of Block I/O Interface}
\label{sec:study}
\begin{figure}[t]
    \centering
    \begin{subfigure}{0.9\linewidth}
        \includegraphics[width=\linewidth]{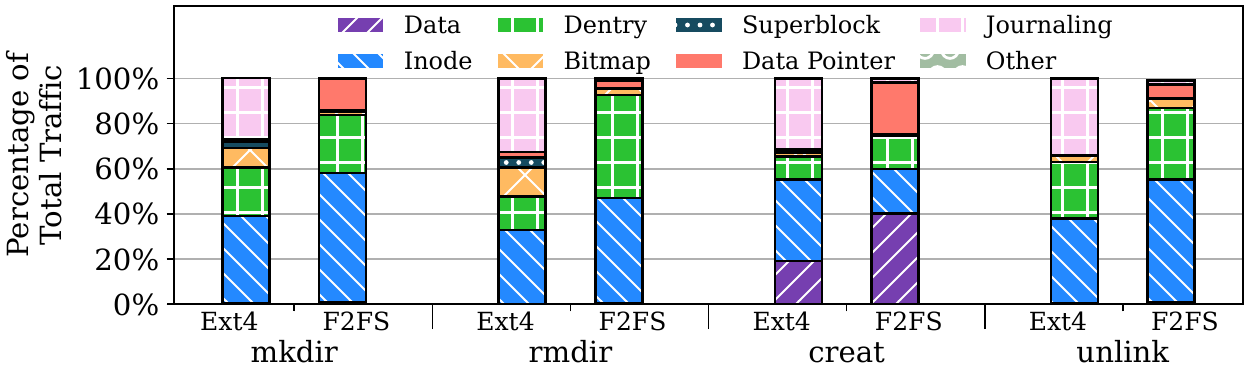}
        \vspace{-3ex}
        \caption{Write traffic breakdown on micro-benchmarks.}
        \label{fig:micro_write_breakdown}
    \end{subfigure}
    \\
    \begin{subfigure}{0.9\linewidth}
        \includegraphics[width=\linewidth]{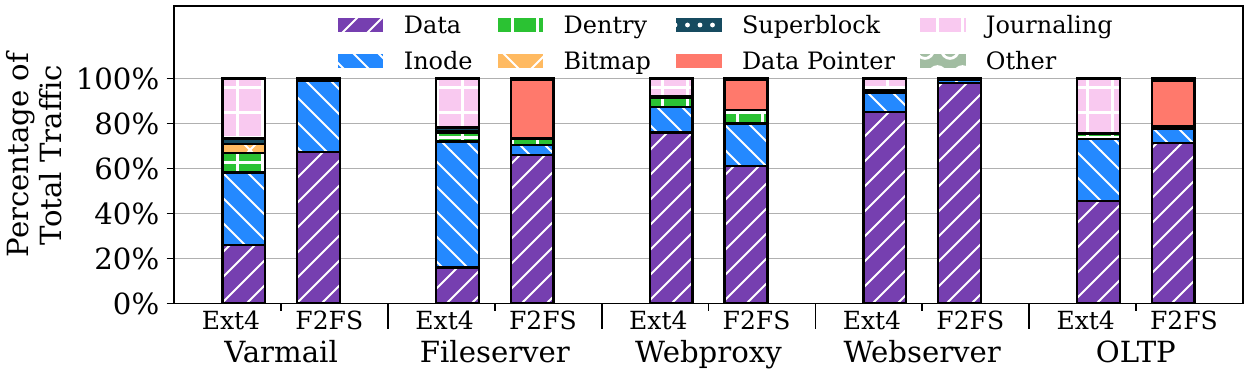}
        \vspace{-3ex}
        \caption{Write traffic breakdown on real applications.}
        \label{fig:macro_write_breakdown}
    \end{subfigure}

    \begin{subfigure}{0.9\linewidth}
        \includegraphics[width=\linewidth]{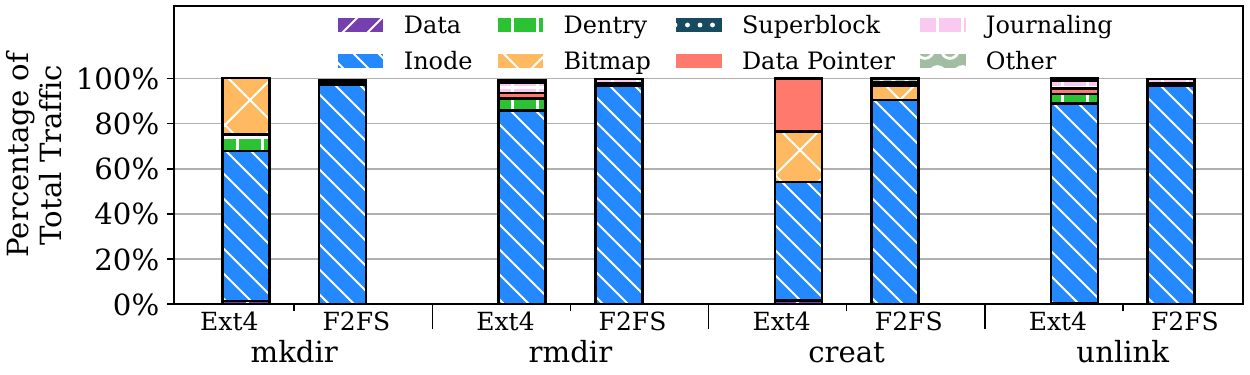}
        \vspace{-3ex}
        \caption{Read traffic breakdown on micro-benchmarks.}
        \label{fig:micro_read_breakdown}
    \end{subfigure}
    \\
    \begin{subfigure}{0.9\linewidth}
        \includegraphics[width=\linewidth]{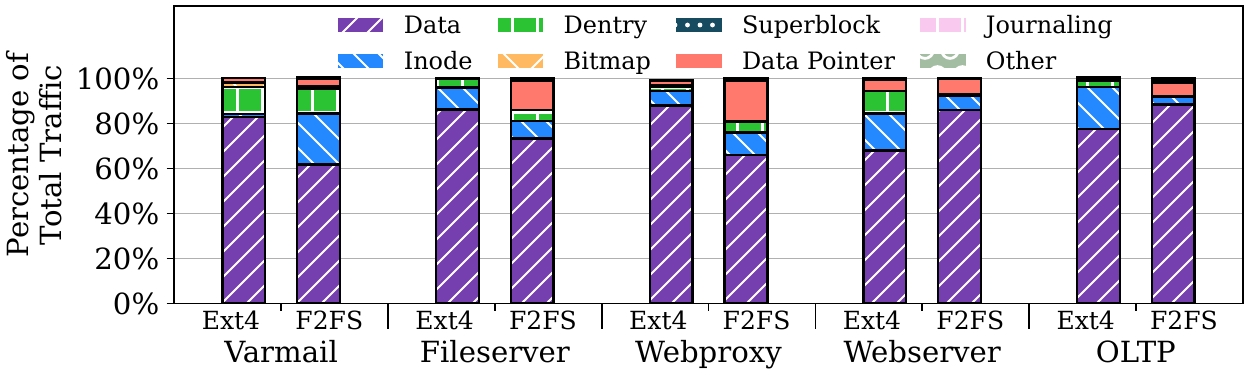}
        \vspace{-3ex}
        \caption{Read traffic breakdown on real applications.}
        \label{fig:macro_read_breakdown}
    \end{subfigure}
    \caption{\hlA{Host-SSD I/O traffic breakdown of Ext4 and F2FS.}}
    \label{fig:motivation_breakdown}
   \vspace{-3ex}
\end{figure}

Before designing \pname{}, \hlB{to further understand the limitations of the block I/O interface and to provide guidelines for the file system design under the block/byte interface}, we conduct a thorough study of the generic Linux file system Ext4 and F2FS. We profile the I/O traffic of the two file systems running Filebench~\cite{tarasov2016filebench} and OLTP~\cite{DifallahPCC13} benchmarks on the server machine described in \S\ref{subsec:eva_setup}. Unlike previous studies on file system I/O~\cite{io:apsys2017}, we focus on the impact of each individual filesystem data structure using the block interface. 
\hlA{We summarize the key data structures in \mbox{Table~\ref{tab:fsops}} and show the traffic breakdown between host and SSD in \mbox{Figure~\ref{fig:motivation_breakdown}}. }

\subsection{I/O Amplification in File Systems}
\label{subsec:io_amplification}
We first focus on the read/write amplification, \hlC{measured by the total read/write traffic to the M-SSD (including both file system metadata and file data) over the read/write traffic issued from the application.}
We collect the I/O amplification factor when running diverse workloads in Table \ref{tab:fs_amp}. As expected, file system operations result in significant write amplification (1.1--6.2$\times$), and metadata operations have a large I/O cost in Ext4 and F2FS. Similarly, the read amplification of Ext4 and F2FS is 1.1--1.7$\times$. 
To further understand the impact of each filesystem data structure, 
we separate the data structures into metadata and data, and explain each of them in detail. 

\subsection{File Metadata Structures}
\label{subsec:metadata_struct}
\noindent
\underline{\textbf{Superblock.}} Superblock maintains the key properties of a file system. As shown in Figure~\ref{fig:motivation_breakdown}, the on-disk superblock is rarely accessed in nearly all workloads at runtime, only contributing 0.23\% write traffic and 0.02\% read traffic on average. Due to the low access frequency, we can still use the block I/O interface 
to simplify its management. 

\noindent
\underline{\textbf{Block/Inode List.}} Block or inode lists are responsible for tracking free data blocks and inodes. In Ext4, they are implemented as bitmaps. F2FS~\cite{F2FS} uses the segment information table (SIT) and node address table (NAT) to track them. Our study with various file operations, such as \texttt{mkdir}, \texttt{rmdir}, and \texttt{create}, show that bitmap accesses contribute 5.4\% of write traffic on average and up to 25.2\% of total read traffic in Ext4, due to the frequent inode/block allocations. During block/inode updates, only a few bytes in the bitmaps are flipped. This offers the opportunity to reduce the write traffic with the byte interface. To minimize the read accesses, we can cache the data structure in the host DRAM after loading it from the storage.

\noindent
\underline{\textbf{Inode.}} File systems use the inode to record the information of files and directories. 
The inode traffic contributes to 35\% and 24.4\% of the total writes on average in Ext4 and F2FS, as inodes are heavily involved in all file and directory updates. Persisting one 128B inode update requires a 4KB write to the disk, further amplifying the write traffic. Therefore, the byte interface can greatly reduce the inode write traffic.

For inode reads, loading an entire inode block brings all the inodes to the host, and they will be cached in the host DRAM. They contribute to 82\% of total reads for metadata-intensive workloads. For data-intensive applications, it is reduced to 12.4\%. Disk accesses can be reduced with data caching in Ext4 and F2FS when accessing files within the same inode block. Therefore, we can load inodes with the block interface and exploit the data locality with metadata caching.

\noindent
\underline{\textbf{Directory Entry.}}
Directory entries (dentries) record the child directories and files within a directory. In Figure~\ref{fig:motivation_breakdown}, the micro-benchmarks involve frequent file/directory creation and removal, and dentries contribute to 23\% of write traffic on average.
We also observe frequent dentry writes in \texttt{Varmail} and \texttt{Fileserver}, as they operate on many files. For workloads with infrequent directory operations, such as \texttt{Webserver} and \texttt{OLTP}, the dentry write traffic is negligible. Thus, to reduce write I/O amplification, the byte interface is preferred. 

To look up a file or directory under a specific directory, the file system needs to search among the dentries. On average, 8\% of the read traffic is spent on reading dentries, and the I/O amplification may further increase as we have deeper or wider directories. Loading the entire dentry structure with the block interface can avoid frequent reads from the storage device.

\noindent
\underline{\textbf{Data Pointer.}} Data pointers record the mappings from the file offset to the logical address of the storage device. Data pointers incur up to 26\% of the total write traffic and 16\% of the read traffic on F2FS. This is because F2FS performs out-of-place updates with frequent data pointer updates. 
To look up the target block address of the file data, we can read the whole block of data pointers into the host with the block interface.

In summary, most metadata updates in file systems are small, so the byte interface is suitable. For their reads, we can use the block interface to exploit data locality.


\subsection{File Data Structures}

\noindent
\underline{\textbf{File Data.}}
File systems allow users to directly access and persist data to the storage bypassing the page cache with Direct I/O. For each I/O request, we can determine the access interface based on the request size in the POSIX call. \hlC{When the write size is less than 512B, writing in cachelines offers a lower write latency than persisting an entire 4KB page. Thus, if a large chunk of data (larger equal than 512B) needs to be accessed, we can use the block interface. For small accesses, we use the byte interface. }

\noindent
\underline{\textbf{Page Cache.}}
File systems use page cache to temporarily cache file data. When a page is not presented in DRAM, the entire page is brought into the host with the block interface. 
To reduce write I/O amplification, we can persist hot cachelines within dirty pages through the byte interface. For largely modified pages,  
we can write them back via the block interface.

\noindent
\underline{\textbf{Data Journal.}} File systems usually employ journaling techniques to prevent metadata/data inconsistency. In Ext4, 30.7\% of the total traffic on average is caused by journaling under the ordered mode, as Ext4 performs double writes for critical metadata structures. F2FS manages node and data blocks with a log structure, and has low write amplification. 
We can use both byte and block interfaces depending on the journal data size. 

In summary, file data may prefer both byte and block interfaces depending on the data locality. We should employ a flexible policy for deciding the appropriate interface at runtime. 


\section{Design and Implementation}
\label{sec:design}

Our goal is to develop a file system \pname{} that transparently supports dual byte and block interface for high-performance data access to memory-semantic SSDs. 

\noindent
\textbf{Design Challenges.}
To achieve this, we need to overcome the following challenges:
(1) There exists an access granularity mismatch between the byte interface and internal flash chips in M-SSDs. \mbox{\pname{}} should minimize the I/O amplification across the storage software and hardware stack.
(2) It is unclear how the file system should support both byte/block interfaces. \pname{} should provide the flexibility to exploit the benefits from both interfaces.
(3) \pname{} should enforce data consistency with minimum overhead.
(4) \pname{} should preserve the essential filesystem properties.


%
%
%

\subsection{System Overview}
\label{subsec:sysoverview}

\begin{figure}[t]
    \centering
    \includegraphics[width=0.8\linewidth]{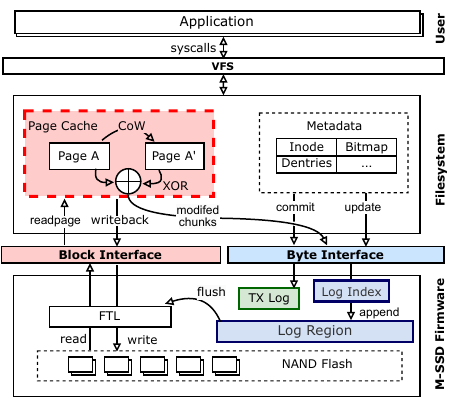}
	\caption{System overview of \pname{}.}
    \label{fig:overview}
\end{figure}

\hlE{We present \mbox{\pname{}}, an efficient file system for M-SSDs with a co-design of software (host filesystem) and hardware (SSD firmware).}
Figure~\ref{fig:overview} shows the system overview of \pname{}.
First, we discuss the techniques for enabling the dual interface in M-SSDs and ensuring persistent writes with the byte interface (\S\ref{subsec:enable_interface}). 
Second, we bridge the gap between byte-granular access and page-granular flash access with a new SSD firmware design. \hlE{We extend the SSD firmware} to manage the SSD DRAM cache as a log-structured write buffer, which enables data coalescing for the flash chip accesses.
We maintain an address mapping to speed up the log lookup and log cleaning process in the SSD firmware (\S\ref{subsec:log}). We then describe how \pname{} supports read and write via the dual interface (\S\ref{subsec:dual_support}).
Third, we \hlE{extend the host-side filesystem metadata and data management} to support the dual interface. \pname{} employs an interface selection mechanism 
based on the data access patterns (\S\ref{subsec:selection} and $\S$\ref{subsec:dataio}).
Fourth, \pname{} ensures crash consistency by enforcing write ordering and atomicity via transactions \hlE{with firmware support.} \pname{} utilizes the in-device write log as a redo log and supports lightweight data consistency and recovery (\S\ref{subsec:consistency}).
Finally, we present a few filesystem operation examples in \pname{} (\S\ref{subsec:put_together}).



\subsection{Enable Byte-granular Data Access/Persistence}
\label{subsec:enable_interface}

The PCIe interface uses Base Address Registers (BARs) to advertise device memory-mappable regions to the host. During system boot-up, the BIOS and OS check the BARs and set up the memory-mapped address space for the device. 
All memory requests to the mapped region are forwarded to the device via the PCIe root complex. The M-SSD controller is responsible for handling memory requests.
\pname{} leverages the BAR register to map the entire SSD as a memory region to the host. \hlE{ The host can concurrently access any SSD address with MMIO (byte interface).
\mbox{\pname{}} do not require cache-coherency between the SSD DRAM and the host CPU cache because the host always contains the latest data and the M-SSD firmware does not modify the data written by the host.}
The byte interface can also be realized with CXL.mem protocol, with which the host CPU can issue cacheable load/store accesses to the device.

To enable persistency, the M-SSD can leverage battery-backed DRAM which allows all data in the DRAM to be flushed to the flash media during a power loss. \hlB{However, modern host CPUs use the Write Combining (WC) mode to reduce PCIe transaction overhead by buffering and coalescing small writes. It may hold dirty cachelines, or a PCIe transaction may be pending upon power loss.}
\pname{} uses two steps to ensure a persistent write. First, it calls \texttt{clflush/clwb} after a memory-mapped write request to flush the CPU cache. Second, it issues a read request with zero byte following the write (write-verify read) to ensure the posted PCIe transactions are completed. Since the read/write requests are serialized in the root complex, a non-posted read will enforce the completion of previous writes.

\begin{figure}[t]
    \centering
    \includegraphics[width=0.85\linewidth]{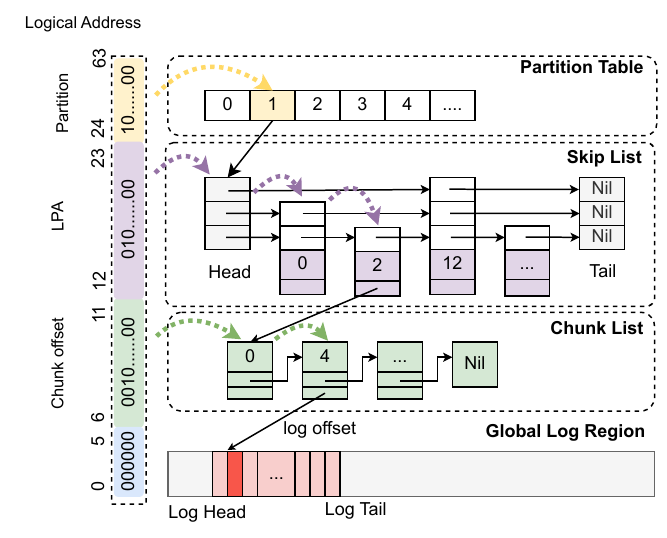}
	\caption{Structure of the write log in the M-SSD firmware.}
    \label{fig:log}
\end{figure}


\subsection{Manage SSD DRAM as a Log-Structured Memory}
\label{subsec:log}

Although PCIe or CXL provides byte-granular access to the SSD, there is still a mismatch of access granularity inside the SSD. The flash chips are only accessible via flash page granularity due to hardware limitations~\cite{gonugondla_iscas, 2bssd, flatflash}, although the host can access the SSD at byte granularity. 
This inevitably incurs extra flash accesses and consumes SSD DRAM cache space.

To bridge the gap between the byte-granular accesses and the page-granular flash accesses, \pname{} reorganizes the SSD DRAM cache into a log-structured region at cacheline granularity. Therefore, all writes via the byte interface can be directly appended to the log to avoid flash accesses on the critical path (see Figure~\ref{fig:overview}). 
\hlC{To exploit data locality and best utilize the SSD DRAM, \mbox{\pname{}} implements a new coordinated caching mechanism between the host and SSD DRAM. Specifically, \mbox{\pname{}} does not have a page-granular cache inside the SSD, and when inevitable flash reads are required, we only cache the loaded pages in the host DRAM, which saves precious SSD DRAM space. }
When the log utilization exceeds a threshold (85\% by default), a background cleaning procedure coalesces the log entries and flushes them back to the flash chips. 

\noindent
\textbf{Index Structure of the Write Log.}
As shown in Figure~\ref{fig:log}, the write log consists of a global log region and an indexing structure to index the log. The global log region buffers the data written via the byte interface. It is organized as a circular buffer (256 MB by default) with head/tail pointers. The written data is appended at the log tail as a 64B-aligned data entry.

We propose an efficient three-layer skip list to index the write log. \hlE{Instead of having a single huge skip list, we break it into multiple smaller ones to minimize the data access latency.} In the first layer, we divide the entire SSD address space (e.g., 1TB) into fixed-size 16MB partitions. During lookup, \pname{} can quickly calculate the partition index by dividing the logical page address (LPA) by the partition size. Then, we find the corresponding skip list from the partition table with the partition index. \hlA{If a single range lookup overlaps several partitions, we break the requests into individual lookups to corresponding partitions.}
In the second layer, 
if a flash page has its data stored in the global log region, an entry indexed by its LPA will be in the skip list. Each skip list entry points to an ordered chunk list in the third layer. The chunk list is ordered by the block offset in a page for fast lookup. Each chunk entry records the block offset (1B) in a page, log offset (4B) in the log region, and data length (4B) of the data entry stored in the global log region.  

The skip list offers $O(log (n))$ lookup, insertion, and deletion time, where $n$ is the number of entries in the skip list. 
Compared to hash tables, our indexing structure supports arbitrary chunk sizes and better range lookup performance. 
\hlE{Since file systems may issue accesses of various sizes, using an arbitrary chunk size does not require breaking larger I/O requests into smaller fixed-size ones. 
} Our experiments with an embedded ARM processor show that the average lookup latency of a fully utilized 256MB log is 89 ns. The overhead is less of a concern, as the flash access latency is at the microsecond scale.

\hlA{The space cost of the log indexing structure solely depends on the log region size and data locality. The log index occupies 21MB of SSD DRAM space on average in our experiment with a 256MB log region. 
}



\noindent
\textbf{Transaction Support with Firmware-Level Logging.}
In \pname{}, all metadata updates are appended to the log region. \hlC{Therefore, we can avoid writing to flash memory twice (double logging) by directly using the write log as a redo log for metadata updates.} This enables \pname{} to support atomic and crash-consistent updates. 

\begin{figure}[t]
    \includegraphics[width=0.8\linewidth]{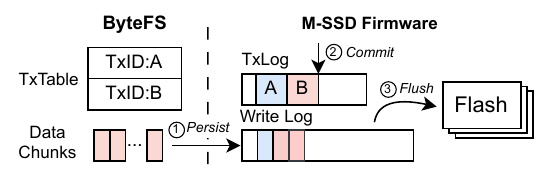}
    \caption{Transaction support with the firmware-level log.}
    \label{fig:consistency}
   
\end{figure}

We show the transaction mechanism of \pname{} in Figure~\ref{fig:consistency}. When \pname{} initiates a new transaction, it assigns a unique 4B transaction ID (TxID) from a monotonically increasing global counter. \pname{} maintains a global transaction table (TxTable) to track all ongoing transactions with their TxIDs. Updates in a transaction are then encoded with the TxID and persisted in the write log (\circleblank{1} in Figure~\ref{fig:consistency}). If the updates from a single transaction exceed the current log space, \mbox{\pname{}} splits the operation into smaller atomic operations at block granularity, which aligns with the atomicity guarantee for operations of existing file systems like EXT4~\mbox{\cite{crash_consistency_osdi14}}. When persisting data in an on-the-fly transaction, concurrent transactions with write requests issued to the same data will be blocked by the transaction-level lock maintained at the host TxTable. When all updates are persisted, \pname{} performs a commit by issuing a custom NVMe command \texttt{COMMIT(TxID)} with the target TxID. To maintain the commit order and status in firmware, we employ a 2MB transaction log (TxLog) in the SSD DRAM. Upon a \texttt{COMMIT(TxID)}, M-SSD firmware appends a 4B commit entry with TxID into TxLog (\circleblank{2}). After the transaction is committed, the M-SSD firmware will be responsible for propagating the update back to
the flash media via log cleaning.
We maintain consistency by flushing the committed updates based on the commit order in TxLog (\circleblank{3}). TxLog is cleaned up after the log cleaning. 




\noindent
\textbf{Log Cleaning.} 
Log cleaning serves to release the space of the log region and persist the updates back to flash chips. We first locate all modified pages by iterating through the second-layer skip lists (line 2 in Algorithm~\ref{alg:log_op}). For each modified page, we reserve a write buffer slot and check whether the old flash page should be loaded in case of partial updates (lines 3-5). We then traverse all modified blocks recorded in the third-layer chunk list and merge the latest committed version into the buffer (lines 6-11). \hlC{If the TxID of the newest updated entry does not appear in TXLog, the entry is not yet committed and we traverse the log region backward to find the latest committed version.} \hlE{All entries that have not been committed are migrated to a new log region.} \hlC{To fully utilize the channel-level parallelism, we group the writes into a large chunk of data in the write buffer and write the data to the flash chips if buffer is filled up (lines 12-13).} Finally, we clean up the log region and the corresponding log indexing structure (line 14).

To prevent a long cleaning process from blocking normal IO accesses, we employ double buffering to serve ongoing writes in a new log region, and flush the old log in the background.

\setlength{\textfloatsep}{2ex}
\begin{algorithm}[t!]
    \DontPrintSemicolon 
    \footnotesize
    \SetAlgoNoEnd
    \SetAlgoLined
    
    
    \SetKwFunction{FMain}{LogFlush}
    \SetKwProg{Fn}{Function}{:}{}
    \Fn{\FMain{list\_head}}{
        {\For{node \textbf{in} \texttt{Traverse}(list\_head)}{
             $buf$ = \texttt{AllocWriteBuffer}($node.lpa$); \\
            {\If{\texttt{CheckPartialUpdate}($node$)} 
                { \texttt{ReadFlashPage}($node.lpa$, $buf$) } 
            }
            {\For{$entry$ \textbf{in} $node.chunk\_list$ }{
                \If{! $entry.committed$} { 
                    \texttt{MigrateToNewLog}($entry$);\\
                    $entry$ = \texttt{GetLatestVersion}(\textit{entry});
                }
                \If{$entry$}{
                    \texttt{memcpy}( \textit{entry.log\_off, buf + entry.off, entry.size});\\
                }
            }
            }  
            {\If{ $WriteBuffer.full$ }
                { \texttt{Buffer\_Write}(); }
            }
        }}
        \texttt{CleanLogRegion}();
    }
    \caption{{Write log cleaning.}} 
    \label{alg:log_op}
    \afterpage{\global\setlength{\textfloatsep}{\oldtextfloatsep}}
\end{algorithm}


\subsection{System Support for Dual Byte/Block Interface}
\label{subsec:dual_support}
We now describe how \pname{} performs read and write accesses with the dual byte and block interface. 

\noindent
\textbf{Read/Write via Byte Interface.} 
\pname{} performs cacheline reads via the mapped SSD memory address space. The host CPU issues memory loads to the M-SSD via PCIe MMIO. For each load, the M-SSD firmware looks up the write log (see $\S$\ref{subsec:log}). If the entry is present in the log, the data is directly returned to the host. Otherwise, the M-SSD fetches the page from the flash chips and only returns the requested cacheline to the host.

\pname{} aligns writes to cachelines as we manage the write log with 64B entries. For a single 64B write, we directly write the cacheline followed by a \texttt{clflush/clwb}.
To write multiple cachelines atomically, we wrap them in a filesystem transaction.
As the data reaches the SSD, the SSD firmware first appends the data into the write log. \hlC{It then inserts the chunk entry for the data to the log index. If the same data is written by two transactions through the byte interface concurrently, we update the chunk entry to track the newest version.}

\noindent
\textbf{Read/Write via Block Interface.}
\pname{} follows the normal block interface to access 4KB blocks with NVMe commands. Upon an NVMe read, the M-SSD firmware loads the page from the flash into a transfer buffer in SSD DRAM. Then, it looks up the skip list using the LPA of the requested flash page. If there are dirty cachelines in the log, the loaded page is merged with the latest data entries. 
Then, the page is returned to the host.

For write requests, 4KB blocks are transferred through PCIe to a write buffer in the FTL layer and then written to flash media. The SSD firmware scans the skip list for the written page and invalidates all corresponding entries in the write log.
These log entries can be invalidated right away since \pname{} ensures that all written-back blocks from the host page cache are up to date.
During later accesses to this page or log flushing, the invalid log entries will be ignored.


\subsection{Manage Metadata Operations in \pname{}}
\label{subsec:selection}

We extend the core metadata structure in \pname{} and \hlE{cache the filesystem metadata in the host DRAM (host-side metadata caching).
Upon cache misses, we load the metadata structure via the block interface.}
We describe the metadata structures and their operations in \pname{} as follows.

\noindent
\textbf{Inodes.} \pname{} maintains the inode as a 128B entry and groups these entries into 4KB pages. To reduce the write traffic of inode updates, we split each inode into the upper and lower regions (64B each). The lower region contains frequently updated information, such as file size, modification times, and access rights. The upper region includes others. Therefore, each inode update takes as low as 64B via the byte interface, which can be done atomically. For complex operations that involve multiple metadata changes, \pname{} utilizes the transaction support (see $\S$\ref{subsec:log}) for atomic updates. 
\pname{} caches inodes using a radix tree in the host memory.
Upon a cache miss, \pname{} loads the entire inode page via the block interface. 


\noindent
\textbf{Directory Entries.} In \pname{}, each directory holds an array of directory entries in its directory blocks. Each entry includes the inode number (4B) of the child file/directory, its file type (2B), filename length (2B), and filename (at most 256B). During directory lookups, we prefer to load the entire directory block once via the block interface, as we need to retrieve the full list of directory entries to read the associated inodes. 
\pname{} caches directory entries by their hashed directory names~\cite{tsai2015get,nova} using a radix tree. 
Creating or renaming a directory involves updating a single directory entry, the update size varies from 64B to 320B based on the filename length, and the byte interface is used to perform the updates.

\noindent
\textbf{Block/Inode Bitmap.} \pname{} maintains the inode and block allocation status with bitmaps. Each bitmap block is divided into multiple 64B groups as the basic unit of update. Upon system boot, \pname{} loads the bitmaps via the block interface. Similar to Ext4, \pname{} keeps a per-CPU free list for scalability and uses an extent-based allocation for data file blocks. Allocating new inodes or blocks incurs frequent small-size bitmap updates, which significantly benefit from the byte interface. 

\noindent
\textbf{Data Pointers.} 
\pname{} uses Ext4-like extent structure to index a range of contiguous file blocks with small extent nodes~\cite{mathur2007new}. Each leaf extent node (16B) includes the file offset (8B), logical block address (4B), and length (4B). When reading or overwriting file data, \pname{} searches the extent tree to find disk pointers, and loads the entire block that contains all extent nodes.
Since \pname{} uses in-place updates for file data, frequent overwrites to data pointers are less common. But it uses byte-granular persistent writes for the updates.
\vspace{-1ex}
\subsection{Manage Data I/O in \pname{}}
\label{subsec:dataio}
\pname{} supports multiple file data I/O modes as normal file systems do, including direct I/O mode, buffered I/O mode, and memory-mapped I/O mode, as well as data journaling. 

\noindent
{\textbf{Direct Data Access.}} 
When users open a file with \texttt{O\_DIRECT} flag, the read and write POSIX calls are performed under the Direct mode. 
When the data access size is no greater than 512B, \pname{} directly accesses the data via the byte interface. 
Otherwise, \pname{} uses the block interface to handle requests like conventional file systems.
This is because when the write size is less than 512B, writing in cachelines offers a lower write latency than persisting an entire 4KB page.  

\begin{figure*}[th!]
    \centering
    \includegraphics[width=0.85\linewidth]{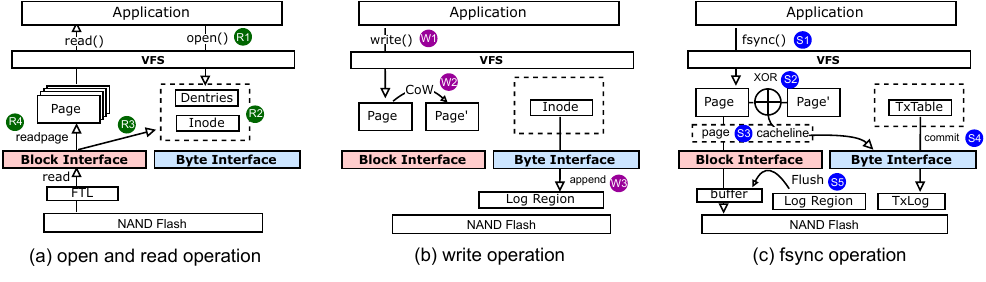}
	\caption{The workflow of \pname{} operations.}
    \label{fig:workflow}
\end{figure*}

\noindent
\textbf{Buffered I/O.}
In buffered I/O mode, the reads/writes are absorbed by the host page cache, and the data access size does not reflect the I/O traffic between the host and SSD. To select between the byte and block interface upon a dirty page writeback, \pname{} calculates the modified ratio ($R$) of this page. 

To obtain the modified ratio, \pname{} uses the copy-on-write (CoW) mechanism in the page cache. When a cached data page is modified, \pname{} copies the original page to a \textit{duplicate page}. Similar to a normal cached page, we track all duplicated pages with an \texttt{XArray}~\cite{XArray} structure indexed by file offset and store them in a per-inode \texttt{struct address\_space}, where all cacheable and mappable objects are tracked.

Upon a dirty page writeback (e.g., triggered by \texttt{fsync}/\texttt{msync}, \hlC{by flush daemon periodically, or under low memory utilization}), \pname{} checks the modified 64B data chunks by XORing the original page and the duplicated page. \pname{} calculates the modified ratio by $R=\frac{N_{\textit{Modified}}}{N_{\textit{total}}}$ where $N_{\textit{Modified}}$ is the number of dirty cachelines and $N_{\textit{total}}$ is the total number of cachelines in the page. If $R$ is less than $\frac 1 8$ (512B of a 4KB page), we use the byte interface to persist small accesses and vice versa. 
\pname{} uses \texttt{avx2}~\cite{avx} instructions available on a majority of CPUs to perform 256-bit XOR operations. Based on our experiments on an Intel server CPU (see $\S$\ref{subsec:eva_setup}), 
an XOR operation takes 936 CPU cycles on average and can achieve up to 14 GB/s throughput. \hlcommon{The CoW mechanism would require extra pages allocated to hold the copied pages. As evaluated with our workloads, the duplicated pages occupy 16\% of the entire page cache size on average, given a maximum page cache size of 8GB. This is less of a concern, considering the significant performance benefits brought by \mbox{\pname{}}. 
 }






\noindent
{\textbf{Memory-Mapped I/O.}}
\hlA{In \mbox{\pname{}}, M-SSD can be used as a memory expander via memory mapping.} \pname{} allows the user-space program to directly access the file data with MMIO via \texttt{mmap()}. To map file data to a specific memory region, \pname{} leverages the cached DRAM pages and maps it to user address space. The interface selection mechanism with buffered I/O can be applied by performing CoW on the mapped file pages.


\noindent
{\textbf{Data Journaling.}} 
\pname{} provides metadata logging that offers the same crash consistency guarantee as the \texttt{ordered} mode in Ext4. In addition, it also provides data journaling. 
For small data writes via the byte interface, we follow the same transaction mechanism in \S\ref{subsec:log} to ensure crash consistency. For large ones written via the block interface, \pname{} combines JBD2~\cite{jbd2}
with \pname{} transactions to support data journaling. Within a transaction, JBD2 writes updated data blocks in a reserved on-disk journal area. \hlC{When commit, \mbox{\pname{}} appends a commit entry containing the corresponding TxID at the end of the JBD2 journal record to identify the commit status for data recovery.}


\subsection{Preserve Essential File System Properties}
\label{subsec:consistency}
\pname{} preserves the essential properties of most file systems, including crash consistency and data recovery.

\noindent
\textbf{Crash Consistency.} 
As discussed in \S\ref{subsec:log}, to ensure consistency, \pname{} leverages transactions to enforce atomicity and write ordering of metadata and data updates. With the firmware-level log-structured memory and battery-backed DRAM in M-SSD, \pname{} accelerates the transaction commits. 




\noindent
\textbf{Data Recovery.} 
M-SSD firmware can retain the cached content in the battery-backed SSD DRAM upon system failures. 
\pname{} can issue a custom NVMe command \texttt{RECOVER()} to the M-SSD firmware to start recovery. 
The M-SSD firmware will perform a complete scan to check all data entries in the log region. For each entry, the firmware checks the 4B TxID encoded at the end of the entry. If the TxID does not appear in the TxLog, we discard it as it is uncommitted. 
Then, the firmware performs a log flush writing all remaining committed entries back to the flash chips based on the order of the TxID in the TxLog. After the firmware finishes the recovery by cleaning up the log region and TxLog, \pname{} performs recovery on data journals if enabled. It scans the journal area for all transactions with a commit entry and moves the data blocks back in place.

\subsection{Put It All Together} 
\label{subsec:put_together}
We show the workflow of \pname{} operations in Figure~\ref{fig:workflow}. We walk through examples of opening and reading a 16KB file, overwriting the first 1KB file chunks, and issuing an \texttt{fsync}.

\noindent\textbf{Open/Read.} An application first \texttt{open}s the target file with the file path (\circlemr{R1}). \pname{} then performs a file lookup on the tree structure (\circlemr{R2}). If a component is not cached in DRAM, \pname{} uses the block interface to fetch the entire block (\circlemr{R3}). After finding the inode and opening the file, the application then issues a 16KB read based on the data pointer. File data is read in blocks and cached in the host page cache (\circlemr{R4}).

\noindent\textbf{Write.} 
The application then issues \texttt{write} syscall to overwrite the first 1KB file data (\circlemw{W1}). \pname{} checks the inode address\_space~\cite{linux:addressspace} to find cached pages, and a copy-on-write is performed on the page (\circlemw{W2}). 
\pname{} allocates a transaction with TxID$_a$ for atomic updates. 
The inode is persisted through the byte interface to the firmware-level write log (\circlemw{W3}).

\noindent\textbf{fsync.} When an \texttt{fsync} is issued (\circlems{S1}), \pname{} first XORs original page and copied page to identify the modified 1KB data chunk (\circlems{S2}). Based on the selection policy, the block interface is selected for the writeback (\circlems{S3}). Since we already persist the metadata updates, we commit the transaction with \texttt{TxCommit}(TxID$_a$) after block I/O is completed (\circlems{S4}). We will flush the committed data to flash chips in the background (\circlems{S5}).

\begin{figure*}[t]
    \centering
    \includegraphics[width=1\linewidth]{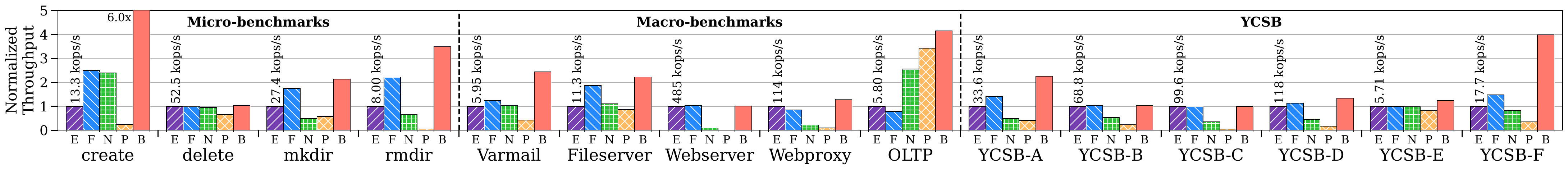}
    \vspace{-4ex}
    \caption{Overall throughput improvement (normalized to Ext4).}
    \label{fig:overall_performance_all}
\end{figure*}

\subsection{Implementation Details}
\label{subsec:impl}

\noindent
\textbf{\pname{} Implementation.} 
We implemented \pname{} as a kernel file system based on Ext4 with 3.9K lines of code (LoC) in Linux kernel 5.1.  \hlB{We reorganize the on-disk metadata structure including superblock, bitmaps, inodes, and directory entries} with 1.3K LoC to support the dual byte/block interface. \hlB{To enable CoW, we add new \texttt{XArray} for indexing duplicate pages and modified the \texttt{writepage()} operation within the file system address space object}~\cite{adress_space} with 0.6K LoC.

\noindent
\textbf{M-SSD Prototype.} To examine the effectiveness of our design, we build a real M-SSD prototype with an OpenSSD FPGA board with 1TB flash storage with 16 channels, and 1GB DRAM. \hlE{The PCIe interface and the flash chip controllers are implemented in the FPGA. The board also contains an onboard ARM core that runs the SSD firmware including FTL functions and DRAM data caching management.} In \mbox{\pname{}}, we modified the SSD firmware with 1.5K lines of C code. It does not require any changes to the SSD hardware. More specifically, we customize the NVMe protocol to enable byte-granularity accesses with 0.1K LoC and implement the 256 MB write log with the three-layer skip list with 0.8K LoC. The log cleaning and transaction support are implemented with 0.4K LoC. We preserve the original SSD FTL layer and its core functionalities, \hlC{in which the address mapping table takes 512MB SSD DRAM and the write buffer takes 16MB SSD DRAM.}

\noindent
\textbf{M-SSD Emulation.}
We also build an M-SSD emulator with 2.1K lines of C code based on the core logic of FEMU \cite{femu}. We dedicate a kernel thread pinned to one CPU core to emulate the normal FTL thread that operates on the embedded SSD processor. We incorporate all core FTL functionalities, such as page allocation, page-level translation, and garbage collection, to emulate the internal structure of M-SSD. To emulate the flash media, we reserve a contiguous region of DRAM memory with \texttt{memmap} boot options and add I/O latency to emulate the NAND flash latency. 
Table~\ref{tab:configuration} shows all the parameters in detail. 






\section{Evaluation} 
\label{sec:eval}
We show that: 
(1) \pname{} outperforms existing file systems by up to 2.7$\times$ when running filesystem benchmarks and real applications (\S\ref{subsec:overall_perf}); 
(2) \pname{} leverages the dual interface to reduce I/O traffic between host and device by up to 5.1$\times$. Its log-structured in-device memory and coalescing mechanism largely reduces flash accesses (\S\ref{subsec:external_traffic}); 
(3) Its individual components are effective (\S\ref{subsec:perf_break});
(4) It can recover from a crash in a short time (\S\ref{subsec:crash}); 
and (5) \pname{} is effective with different device configurations (\S\ref{subsec:sensitivity}).

\begin{table}[t]
    \footnotesize
        \footnotesize
    \caption{System configurations.}
    \vspace{-2ex}
    \centering
    \begin{tabular}{p{51pt} p{33pt} p{74pt} p{40pt}}
       \hline        
    \multicolumn{4}{c}{Host Machine Configuration} \\ \hline
    \multicolumn{2}{l}{Processor} &\multicolumn{2}{l}{Intel(R) Xeon(R) E5-2683 v3}\\
    \multicolumn{2}{l}{Memory Size}& \multicolumn{2}{l}{128 GB}\\
    \hline
       \multicolumn{4}{c}{SSD Emulator Configuration} \\ \hline 
       
     Capacity & 32 GB & R/W Bandwidth & 3.5/2.5 GB/s \\
     Page Size & 4 KB &  Flash R/W Latency  & 40/60 $\mu$s \\
     Channel Count & 8 & Cacheline R/W Latency & 4.8/0.6 $\mu$s \\
     
       \hline

    \end{tabular}
    \label{tab:configuration}
\end{table}

\begin{table}[t]
    \footnotesize
        \footnotesize
    \caption{ Workloads used in our evaluation. }
    \vspace{-2ex}
    \centering
    \begin{tabular}{p{78pt} p{136pt}}
       \hline
        \textbf{Workload} & \textbf{Description} \\ \hline
        
    \multicolumn{2}{c}{Microbenchmarks} \\ \hline
    \texttt{create} & File creation, 1M 4KB files, 12 threads  \\
    
    \texttt{delete} & File deletion, 1M 4KB files, 12 threads \\
    
     \texttt{mkdir} & Make directories, 1M dirs, 12 threads \\

     \texttt{rmdir} & Remove directories, 1M dirs, 12 threads \\
       \hline
       \multicolumn{2}{c}{Applications}\\ \hline 
       \texttt{Varmail} & 1M 16KB files, 12 threads  \\
       \texttt{Fileserver} & 100k 128KB files, 12 threads  \\
       \texttt{Webproxy} & 1M 16KB files, 12 threads  \\
       \texttt{Webserver} & 1M 16KB files, 12 threads  \\
       \texttt{OLTP} & 1.6K 10MB files, 200 threads  \\
    \texttt{YCSB} on RocksDB & 10M \hlC{1000B} KV pairs, 40M ops, zipfian distribution\\\hline
    \end{tabular}
    \label{tab:workloads}
\end{table}

\subsection{Evaluation Setup}
\label{subsec:eva_setup}
We run \pname{} on a dual-socket, 28-core Intel(R) E5 platform with a base frequency of 2.7GHz and 128GB memory. Our SSD platform includes both a real SSD prototype and an SSD emulator as discussed in \S\ref{subsec:impl}. For emulation, we emulated SSD with 4KB flash pages and a log region of 256MB. We list all core settings in Table~\ref{tab:configuration}. 
We use various workloads including file system benchmarks Filebench~\cite{tarasov2016filebench} and real-world applications YCSB on RocksDB~\cite{rocksdb}, as shown in Table~\ref{tab:workloads}. 
We compare \pname{} with state-of-the-art file systems, including Ext4, F2FS, NOVA, and PMFS.  
\hlE{As Ext4 and F2FS are designed for block-based SSDs, we mount them on the M-SSD with the default block interface. To set up NOVA and PMFS that are designed for byte-addressable NVM, we map the entire M-SSD space to the host and mount them on the mapped M-SSD region in the same way as they are mounted on NVM.} \hlcommon{All baseline file systems are running on the M-SSD without firmware changes (i.e., no log-structure memory in SSD DRAM), but we enabled the data caching (256MB) in SSD DRAM. \mbox{\pname{}} uses the 256MB data cache in SSD DRAM as a log-structured memory region. The remaining SSD DRAM space is used to maintain the essential functions (e.g., address mapping table) of the flash translation layer of an SSD.} \hlC{For the SSD prototype, we format the file system for the entire 1TB SSD space, and for the sensitivity experiments with the emulator, the file system is 32GB.} In figures, Ext4, F2FS, NOVA, PMFS, and ByteFS are shown as `E', `F', `N', `P', and `B'.


\subsection{Overall Performance Improvements}
\label{subsec:overall_perf}
We evaluate \pname{} with filebench micro-benchmarks, macro-benchmarks, and YCSB workloads on the real SSD prototype. We show the throughput improvement in Figure~\ref{fig:overall_performance_all}.

\noindent
\textbf{Micro-benchmarks.} 
\pname{} outperforms Ext4 by 2.5$\times$ and F2FS by 1.48$\times$ on average across all micro-benchmarks. On file creation especially, \pname{} achieves 6.0$\times$ and 2.4$\times$ performance improvement compared to Ext4 and F2FS. \pname{} can also obtain 1.2$\times$-1.5$\times$ performance speedup on \texttt{mkdir} and \texttt{rmdir} when compared to F2FS, because \pname{} involves fewer data movements between host and device to perform file system operations with the dual interface. On \texttt{delete}, \pname{} can achieve similar performance with Ext4 and F2FS, since delete (i.e., unlink) operations do not require an immediate sync to the device. NOVA and PMFS perform even worse than EXT4 and F2FS in most cases since they are not designed for flash-based SSDs, they purely rely on the byte interface which fails to exploit the spatial locality with the block interface.

\begin{figure*}[th]
    \centering
    \begin{subfigure}{0.47\linewidth}
        \includegraphics[width=\linewidth]{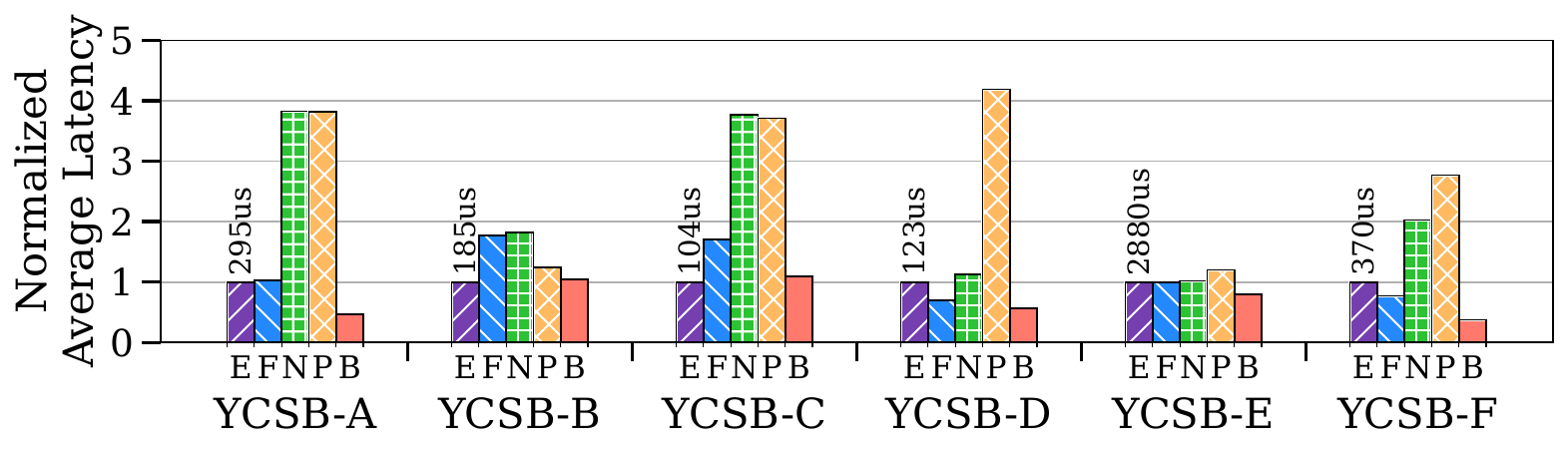}
        \vspace{-4ex}
        \caption{Read Average Latency}
        \label{fig:read_avg_latency_ycsb}
    \end{subfigure}
    \hfill
    \vspace{-0.5ex}
    \begin{subfigure}{0.47\linewidth}
        \includegraphics[width=\linewidth]{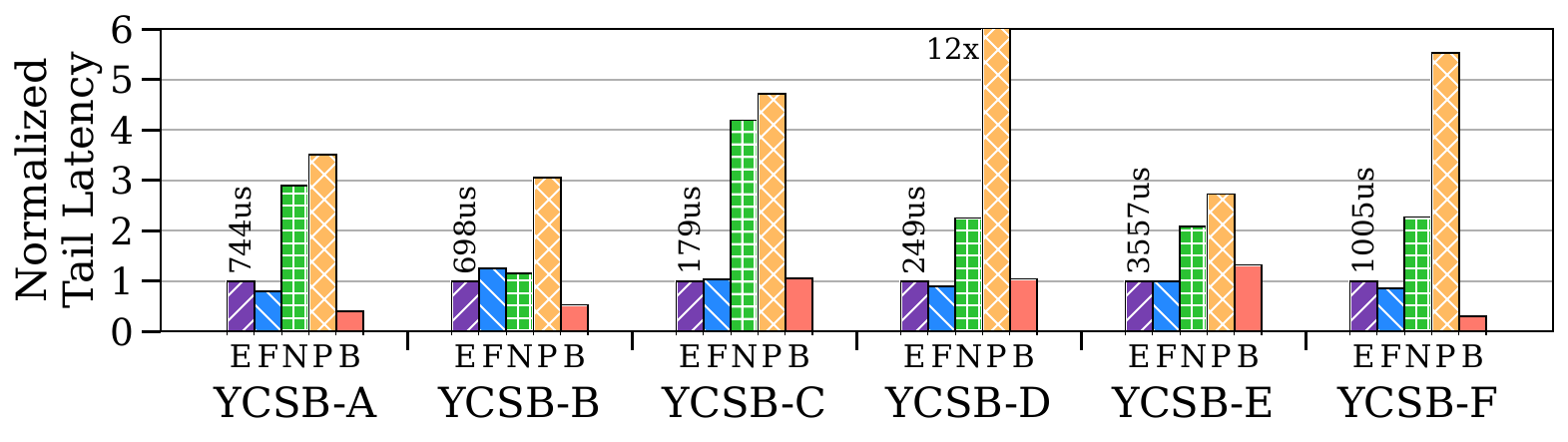}
        \vspace{-4ex}
        \caption{Read 95\% Tail Latency}
        \label{fig:read_95_latency_ycsb}
    \end{subfigure}
    \begin{subfigure}{0.47\linewidth}
        \includegraphics[width=\linewidth]{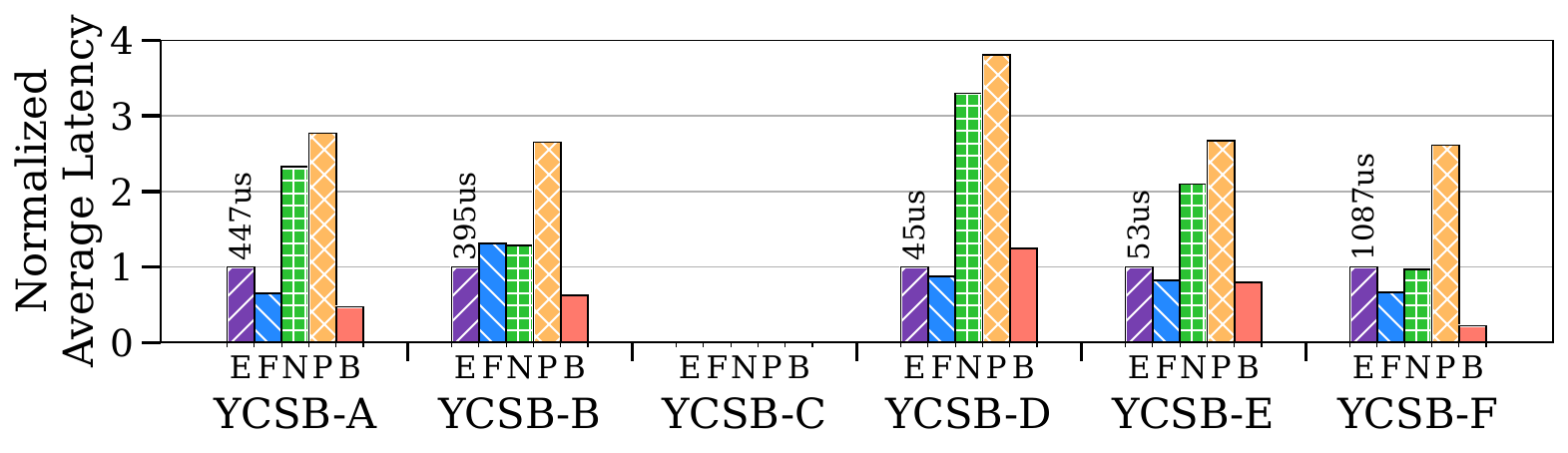}
        \vspace{-4ex}
        \caption{Write Average Latency}
        \label{fig:update_avg_latency_ycsb}
    \end{subfigure}
    \hfill
    \begin{subfigure}{0.47\linewidth}
        \includegraphics[width=\linewidth]{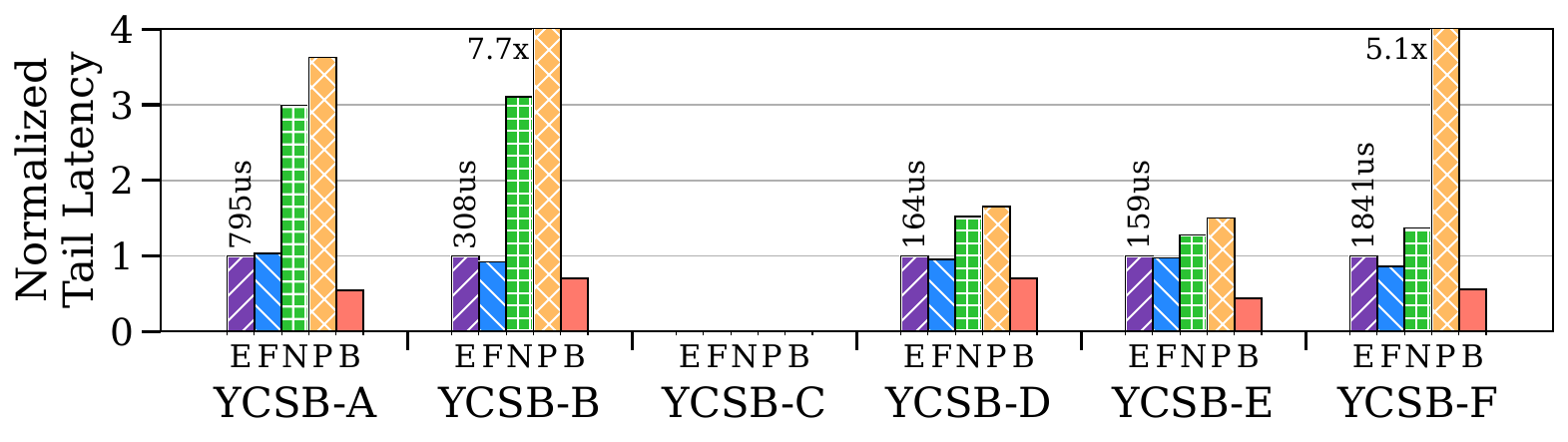}
        \vspace{-4ex}
        \caption{Write 95\% Tail Latency}
        \label{fig:update_95_latency_ycsb}
    \end{subfigure}
    \caption{Latency of YCSB workloads (normalized to Ext4). YCSB-C does not have write latency as it is read-only.}
    \label{fig:ycsb_latency}
\end{figure*}




\noindent
\textbf{Macro-benchmarks.} \pname{} achieves a higher or similar throughput across all macro-benchmarks. For \texttt{Varmail}, \pname{} outperforms F2FS by 1.9 $\times$, as \texttt{Varmail} frequently creates small files and performs small synchronous I/Os. \pname{} can speed up the frequently issued syscalls including \texttt{create} and \texttt{fsync} by persisting small metadata via the byte interface.

\texttt{Fileserver} is a data-intensive workload that reads and appends to relatively large files. \pname{} uses block interface to persist large amounts of data to the disk with high parallelism, while reducing the critical metadata writes caused by frequent appending with byte interfaces. As a result, \pname{} achieves over 2.2$\times$ throughput improvement compared to Ext4, and 1.2$\times$ improvement compared to F2FS.

\texttt{Webserver} and \texttt{Webproxy} are read-heavy workloads. \pname{} shows a similar performance compared with Ext4 and F2FS, as we leverage the block interface and host-side caching to exploit data locality. \pname{} outperforms Ext4 by 1.3$\times$ on \texttt{Webproxy}, since \texttt{Webproxy} also involves heavy directory operations, which benefits from the byte interface in \pname{}.

Finally, \pname{} outperforms Ext4 by 4.1$\times$ on write-intensive workload \texttt{OLTP}. It creates over 200 threads creating, appending, and overwriting files with frequent \texttt{fdatasync}. NOVA and PMFS can use byte interface to directly persist small updates, which reduces sync overhead and scales better under high contention. \pname{} further outperforms them by reducing consistency overhead with the in-device write log. 


\noindent
\textbf{YCSB Workloads.} 
To evaluate \pname{} on real-world applications, we run YCSB workloads on RocksDB. 
We report \hlD{the end-to-end} average latency and 95\% tail latency for Read and Update in Figure~\ref{fig:ycsb_latency}. 
\pname{} provides 2.4$\times$ better throughput compared to F2FS. Similarly, \pname{} offers lower average and tail latency.
Its major benefit comes from reducing the critical-path write latency, which also improves read latency as the write requests may block the read requests in RocksDB.
For \texttt{YCSB-A} and \texttt{YCSB-F} which has a read/write ratio of 50/50, \pname{}
improves the average/tail latency by 2.3$\times$/2.0$\times$ for reads and 1.3$\times$/1.6$\times$ for writes compared to F2FS.
In workloads with a lower write ratio (i.e., \texttt{YCSB-B} and \texttt{YCSB-D} with a 95/5 read/update ratio), \pname{} provides relatively less performance improvement over EXT4.
In \texttt{YCSB-C} with 100\% read operations, \pname{} 
achieves similar performance as EXT4 or F2FS.
It also achieves similar performance on \texttt{YCSB-E} compared to other baselines, since it performs range scanning following a uniform distribution on the entire dataset, and has almost no locality.

\begin{figure}[t]
    \centering
    \includegraphics[width=0.95\linewidth]{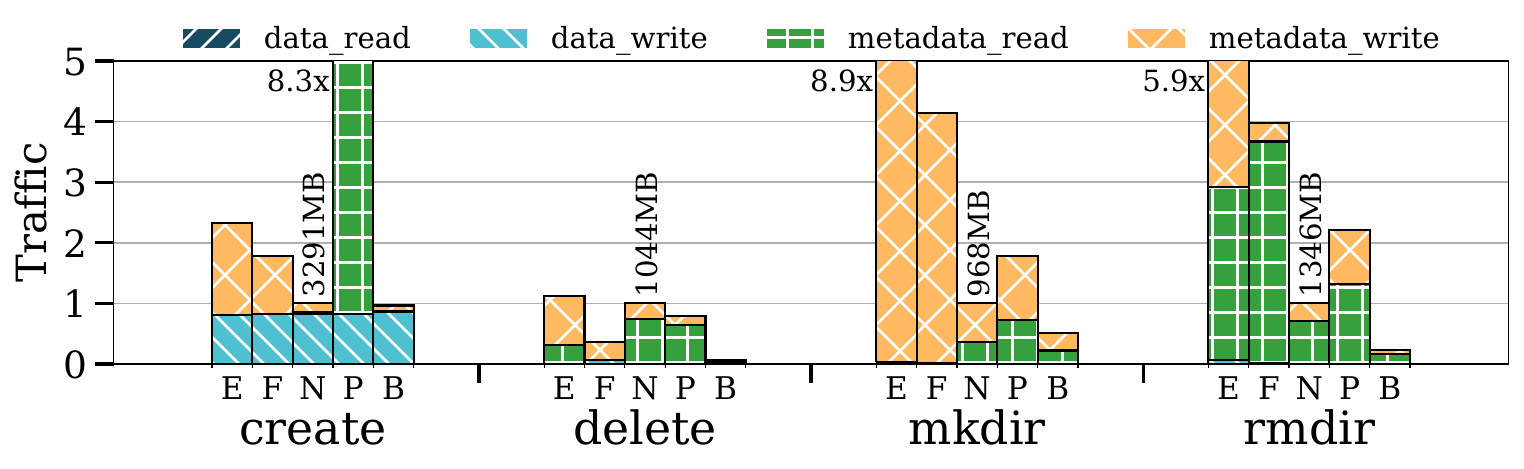}
    \vspace{-2ex}
    \caption{Host-SSD I/O traffic breakdown with Filebench Micro-benchmarks (normalized to NOVA).}
    \label{fig:external_traffic_micro}
\end{figure}
\begin{figure}[t]
    \centering
    \includegraphics[width=0.95\linewidth]{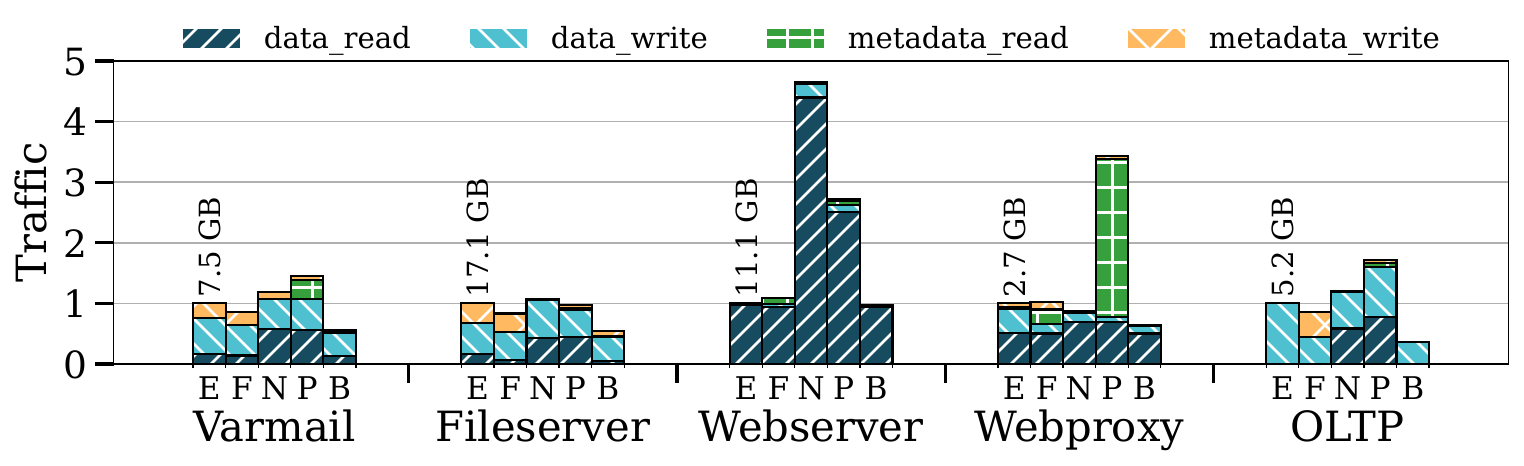}
    \vspace{-2ex}
    \caption{Host-SSD I/O traffic breakdown with Filebench Macro-benchmarks (normalized to Ext4).}
    \label{fig:external_traffic_macro}
\end{figure}

\subsection{I/O Traffic Breakdown}
\label{subsec:external_traffic}

\textbf{I/O Traffic Between Host and SSD.}
Figure~\ref{fig:external_traffic_micro} shows the traffic between host and SSD with micro-benchmarks.
Compared to Ext4 and F2FS which only use block interface, \pname{} reduces the metadata traffic by up to 25.3$\times$ and 17.2$\times$ (11.4$\times$ and 6.1$\times$ on average) with the byte interface. Compared to NOVA or PMFS, which also use the byte interface, \pname{} can still reduce the metadata traffic.
This is for two reasons. First, NOVA uses out-of-place updates and \hlE{PMFS uses data journaling} to ensure crash consistency, while \pname{} makes in-place updates of metadata and reduces consistency overheads with the write log in the SSD firmware, thus avoiding the double writes on the metadata. Second, \pname{} reduces metadata reads with block interface and host-side metadata caching.
Compared to NOVA, \pname{} reduces nearly 43\% metadata read traffic on average.
\mbox{\pname{}} also reduces the metadata read traffic compared to Ext4 and F2FS. For some metadata writes (e.g., in \texttt{delete}/\texttt{rmdir}), if the corresponding data is not cached in the host, Ext4 needs to load the entire block, even though only part of the block needs to be updated. \mbox{\pname{}} can directly update the requested data at byte granularity, reducing read traffic.

Figure~\ref{fig:external_traffic_macro} shows the I/O traffic for the macro-benchmarks.
Similar to the micro-benchmarks, \pname{} significantly reduces the metadata traffic. 
In addition, \pname{} also reduces data traffic with the dynamic block/byte interface selection and host-side page caching.
\texttt{Varmail} and \texttt{Fileserver} combine frequent file creation and deletion operations with heavy file append operations. Compared to Ext4 and F2FS, \pname{} reduces the metadata update traffic with the byte interface. Compared to NOVA and PMFS, \pname{} reduces data read overhead by adaptively exploiting the block interface.
In read-heavy workloads \texttt{Webserver} and \texttt{Webproxy}, \pname{} reduces the data traffic by up to 4.7$\times$ and 2.7$\times$ compared to NOVA and PMFS. \texttt{Webproxy} also involves heavy directory operations, so \pname{} reduces metadata traffic compared to Ext4 and F2FS. In \texttt{OLTP}, \pname{} reduces the data write traffic by 1.6$\times$ and 2.2$\times$ compared to NOVA and PMFS. Both of them incur extra write traffic due to their page-granular copy-on-write mechanism to maintain crash consistency, while \pname{} employs the byte-granular in-device write log.

%
  
\begin{figure}[t] 
    \centering
    \includegraphics[width=0.95\linewidth]{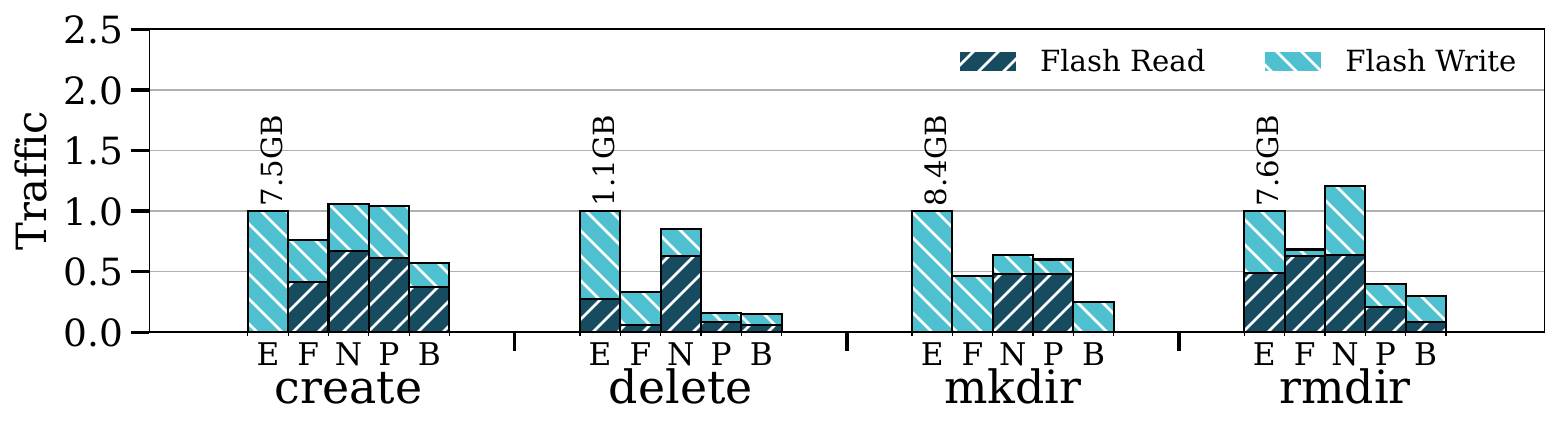}
    \vspace{-1ex}
    \caption{SSD flash traffic breakdown with Filebench Micro-benchmarks (normalized to Ext4).}
    \vspace{-1ex}
    \label{fig:internal_traffic_micro}
\end{figure}
\begin{figure}[t]
    \centering
    \includegraphics[width=0.95\linewidth]{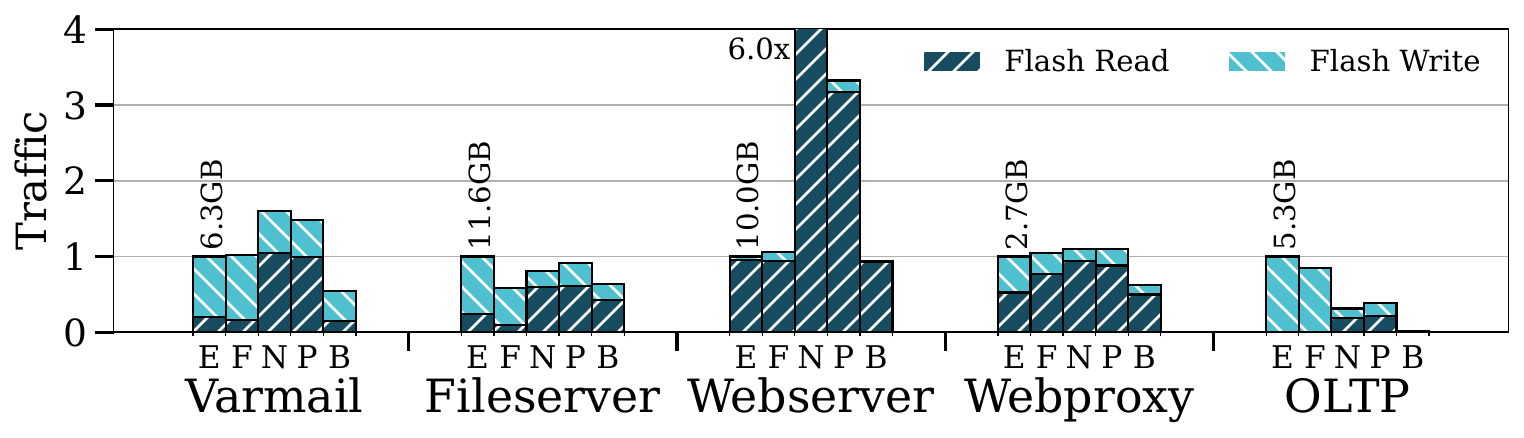}
    \vspace{-1ex}
    \caption{SSD flash traffic breakdown with Filebench Macro-benchmarks (normalized to Ext4).}
    \label{fig:internal_traffic_macro}
\end{figure}

\noindent
\textbf{SSD Flash Traffic.}
We now evaluate the effectiveness of the log-structured memory in the SSD. Figure~\ref{fig:internal_traffic_micro} and Figure~\ref{fig:internal_traffic_macro} show the flash traffic in the SSD. On average, \pname{} reduces the flash traffic by 2.9$\times$, 2.1$\times$, $3.2\times$, and $2.2\times$ compared to Ext4, F2FS, NOVA, and PMFS.
\pname{} reduces flash write traffic by coalescing small writes in the in-device write log. \pname{} also reduces flash read traffic as it does not need to fetch the corresponding page from the flash upon a partial write.

Sometimes, \pname{} may incur higher flash traffic than other file systems. For example, in \texttt{create} and \texttt{Fileserver}, \pname{} incurs higher flash read traffic than EXT4 and F2FS. 
In \texttt{mkdir} and \texttt{rmdir}, \pname{} may incur higher flash write traffic than NOVA and PMFS.
This is caused by the read-modify-write pattern with partial writes during the log-cleaning stage.
On the other hand, NOVA and PMFS use page-granular cache in the SSD DRAM, so a cached page may absorb more writes when the locality is good enough.
The higher flash traffic in \pname{} does not affect performance since the log cleaning happens in the background. Such overhead is completely outweighed by the performance benefits of the in-device write log.




\begin{figure}[t!]
    \centering
    \includegraphics[width=0.95\linewidth]{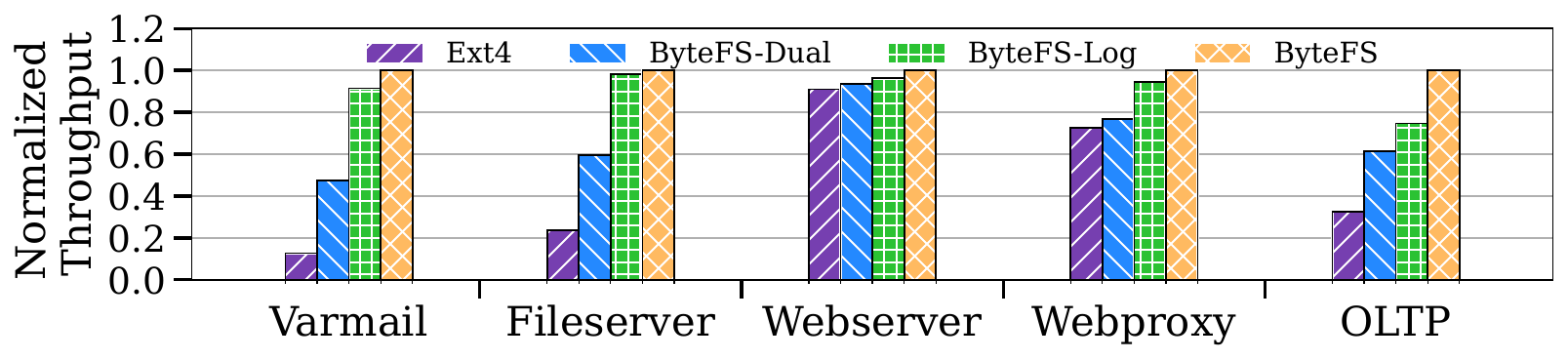}
    \vspace{-1ex}
    \caption{\pname{} perforamce breakdown.  \textit{ByteFS-Dual}: ByteFS with only dual interface. \textit{ByteFS-Log}: ByteFS-Dual with log-structured memory. \textit{ByteFS}: the full ByteFS design.}
    \label{fig:perf_break}
   \vspace{-1ex}
\end{figure}

\vspace{-1ex}
\subsection{Performance Breakdown of \pname{}}
\label{subsec:perf_break}

To understand how each design component in \pname{} contributes to the overall performance, we evaluate three variants of \pname{} in Figure~\ref{fig:perf_break}. 
\hlC{In ByteFS-Dual, we enable dual byte/block interface for metadata operations, but for data operations, we still use the block interface. And M-SSD still uses page-granular caching. ByteFS-Log is similar to ByteFS-Dual. The only difference is that the firmware-level logging is enabled in M-SSD. In ByteFS, we enable dual byte/block interface for both metadata and data operations. And the firmware-level log-structured memory region is enabled in M-SSD.} 
\texttt{Varmail} and \texttt{Fileserver} benefit from both the dual interface and the log-structured write buffer, as they help reduce the I/O amplification of these two workloads. \texttt{Webproxy} mostly benefits from the dual interface, as it involves heavy directory operations. \texttt{OLTP} benefits from both log-structured memory and flexible interface selection, as it involves frequent small overwrites. 

\vspace{-1ex}
\subsection{Recovery Time of \pname{}}
\label{subsec:crash}

To measure the recovery time of \pname{} after a system crash, we intentionally power off the system after running YCSB workloads. \pname{} can recover the system in 4.2 seconds on average after reboot, with the recovery process described in \S\ref{subsec:consistency}. It takes 0.9 seconds to load the entire SSD DRAM content and 2.7 seconds to scan the global log region and TxLog to flush all committed entries into the flash media.






\begin{figure}[t]
    \centering
    \includegraphics[width=\linewidth]{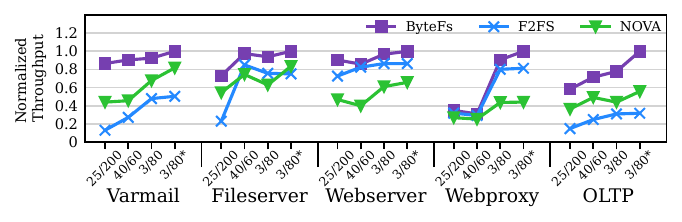}
    \vspace{-4ex}
    \caption{Macro benchmark performance under different flash latency (\texttt{Read/Write}).}
    \label{fig:sensitivity_access}
\end{figure}
\begin{figure}[t]
    \centering
    \includegraphics[width=0.95\linewidth]{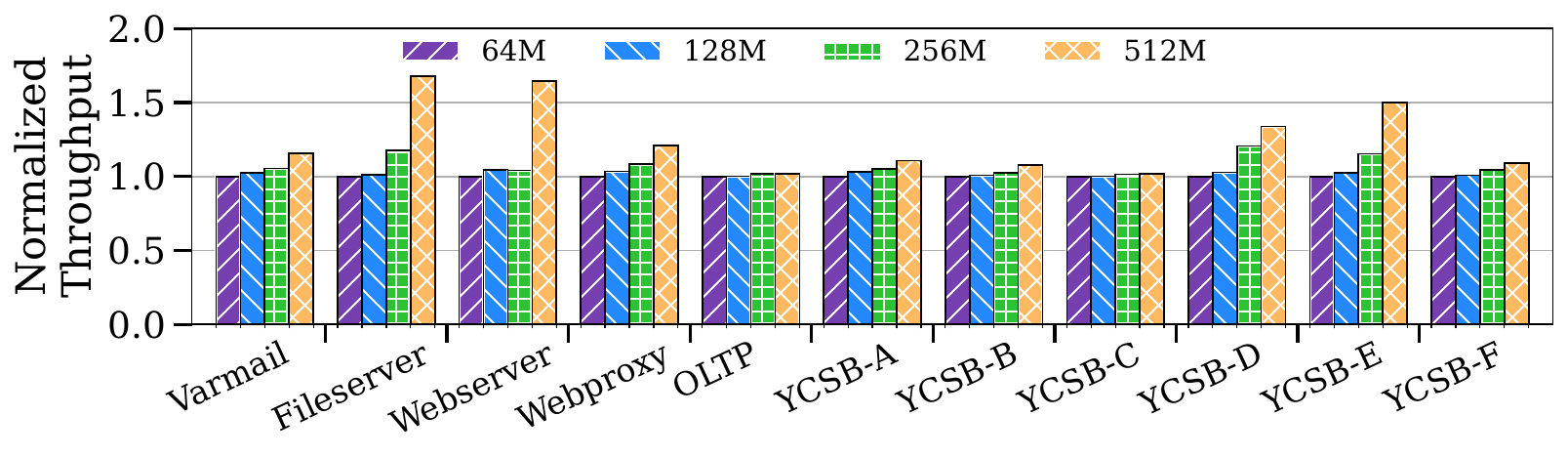}
    \vspace{-2ex}
    \caption{Performance impact of the log region size (normalized to 64MB).}
    \label{fig:Sensitivity_log_size}
\end{figure}

\vspace{-1ex}
\subsection{Sensitivity Analysis}
\label{subsec:sensitivity}
\textbf{Vary M-SSD Access Latency.} We vary the emulator configuration with various NAND flash read/write latencies from low-end to high-end SSDs~\cite{optanessd2,cheong2018flash,atc-cxlssd}. As shown in Figure~\ref{fig:sensitivity_access}, \pname{} outperforms F2FS and NOVA regardless of flash access latencies. We also emulate a CXL-based SSD with less cacheline access latency (175 ns~\cite{cxl_latency}) and high-end flash media (marked as 3/80*).
The benefit of \pname{} is larger with higher flash write latency, as \pname{} hides the flash write latency by flushing the in-device write log in the background. With CXL, both \pname{} and NOVA have better performance in some workloads as the CXL protocol reduces the access latency of the byte interface. However, NOVA is still slower than \pname{} since it does not optimize for the high flash access latency.

\noindent
\textbf{Vary SSD DRAM Log Size.} As shown in Figure~\ref{fig:Sensitivity_log_size}, \pname{} can scale its performance with a larger SSD DRAM. As we increase the write log size, \pname{} can cache and coalesce more updated entries inside the DRAM before flushing. While a larger log size might lead to a longer flush time, many workloads still benefit from a larger log size, as the log flushing is done in the background. For some workloads (e.g., OLTP) the benefits are marginal as they already have good write locality.

\vspace{-1ex}
\section{Related Work}
\label{sec:related}

\textbf{Storage Class Memory.} 
The hardware community has been focused on developing scalable memory technologies such as PCM~\cite{mem_pcm_1}, ReRAM~\cite{mem_reram}, and FeRAM~\cite{www:fram} to increase the memory capacity. 
However, many of them are not available in real products. 
Intel released the Optane persistent memory~\cite{intel:xpoint} in 2019, but it was shut down in 2022. This motivates the community to investigate alternatives. Most recently, research demonstrated the feasibility of memory-semantic SSDs~\cite{2bssd, memorysematicssd}. \pname{} targets this new type of storage device and enables the system software support to exploit its unique properties.

\noindent
\textbf{Memory-Semantic SSDs.}
\hlE{To exploit the byte-addressability of SSDs, FlatFlash\mbox{~\cite{flatflash}}
proposs a software-based implementation of M-SSD by integrating the SSD into a unified virtual memory space via memory mapping, and enabling unified address translation for host DRAM and flash memory.}
With the new CXL protocols, Jung et al. developed an FPGA-based emulator~\cite{jung2022hotstorage} for mapping the SSD as a memory extension. 
However, few of them investigated how the M-SSD should be managed at the systems software side. ByteFS provides efficient system software support for M-SSDs. It rethinks the SSD firmware design to address the fundamental mismatch of data access granularity between the SSD DRAM and flash chips.  
 
\noindent
\textbf{Persistent Memory Systems.}
To support byte-addressable persistent memory, prior studies have developed a variety of file systems~\cite{bpfs, pmfs, nova, splitfs}. For example, 
BPFS~\cite{bpfs} developed an optimized file system in OS kernel and optimized shadow paging technique for ensuring crash consistency. PMFS~\cite{pmfs} optimized the data access to persistent memory via direct access. NOVA~\cite{nova} implemented a per-inode log-structured file system. SplitFs~\cite{splitfs} handled data operations entirely in the user space and processed the metadata operations in Ext4. 
\hlD{Other file systems targeted a tiered memory setting. Specifically, Ziggurat\mbox{~\cite{ziggurat}} and NVMFS\mbox{~\cite{nvmfs}} are monolithic tiered file systems, with a focus on data placement across tiered memory. SPFS\mbox{~\cite{spfs}} and Strata\mbox{~\cite{Starta}} are stackable file systems, which stack persistent memory file systems upon the block-based file systems.}
Unlike these prior studies, \pname{} develops a file system for M-SSDs that have fundamentally different device characteristics. 



\section{Conclusion}
\label{sec:conclusion}
We present \pname{}, a new kernel-level file system for memory-semantic SSDs that support dual byte/block interface. \pname{} significantly reduces the I/O amplification across the entire storage stack with various optimizations in the core components of file systems. 
We implement \pname{} on both a programmable SSD FPGA board and an SSD emulator to demonstrate its efficiency and usability.  

\begin{acks}
We thank the anonymous reviewers and our shepherd Haris Volos for their insightful comments and feedback. 
We thank the members in the Systems Platform Research Group at University of Illinois Urbana-Champaign for constructive discussions. 
We also thank Yuqi Xue for proofreading the paper. This work was partially supported by NSF grants CCF-2107470 and CCF-1919044.
\end{acks}

\bibliographystyle{plain}
\balance
\bibliography{ref,mlref,flashblox-ref,flatflash-ref,nvmlog}

\clearpage
\appendix
%
%
%
%
%



\definecolor{codegreen}{rgb}{0,0.6,0}
\definecolor{codegray}{rgb}{0.5,0.5,0.5}
\definecolor{codepurple}{rgb}{0.58,0,0.82}
\definecolor{backcolour}{rgb}{0.95,0.95,0.92}




\lstset{style=BashStyle}




\section{Artifact Appendix}

\subsection{Abstract}
We implement \pname{} by building our emulation framework and an SSD prototype
described in (\S\ref{subsec:impl}). In this artifact, we provide the source code of \pname{} within the emulator version
and necessary instructions to reproduce key performance results
(Figure 6-11 in \S~\ref{sec:eval}). The emulator can be executed on any x86 machine with at least 96
GB of main memory and at least 64 GB of disk space. We strongly
recommend running the artifact with the emulator on a workstation with multi-cores
and at least 128 GB memory.


\subsection{Artifact check-list (meta-information)}


{\small
\begin{itemize}
  \item {\bf Program:} Filebench, YCSB on RocksDB.
  \item {\bf Compilation:} GCC 7.5.0
  \item {\bf Run-time environment:} Ubuntu 18.04.
  \item {\bf Hardware:} X86 machine with at least 96 GB of main memory and at least 64 GB of disk space for the emulation version. OpenSSD FPGA board for SSD prototype.
  \item {\bf Metrics:} Throughput, latency and I/O traffic.
  \item {\bf Output:} Files and graphs, expected results included in the repo.
  \item {\bf Experiments:} Need kernel compilation and running benchmarks using supplied scripts.
  \item {\bf How much disk space required (approximately)?:} 64 GB
  \item {\bf How much time is needed to prepare workflow (approximately)?:} 1-2 hours including kernel compilation on a 64-core server machine.
  \item {\bf How much time is needed to complete experiments (approximately)?:} 8 hours for all benchmarks.
  \item {\bf Publicly available?:} Yes.
  \item {\bf Archived (provide DOI)?: }10.5281/zenodo.14321357
\end{itemize}
}

\subsection{Description}

\subsubsection{How to access}
The latest source code and instructions can be found at
GitHub repository at \url{https://github.com/platformxlab/bytefs.git}.

\subsubsection{Hardware dependencies}
The artifact with the emulator can be executed on any x86 machine with at least 96 GB of main memory and at least 64 GB of disk space.

\subsubsection{Software dependencies}
The artifact needs a Linux environment (tested on Ubuntu 18.04).



\subsection{Installation}

This part we will describe how to build the artifact with the emulator version. Start by downloading the artifact from the GitHub repository:
\begin{lstlisting}[language=bash]
git clone https://github.com/pollux006/ByteFS_AE.git
cd ByteFS_AE
\end{lstlisting}

Checkout branch according to the version you want to build:
\begin{lstlisting}[language=bash]
git checkout baseline # or
git checkout bytefs
\end{lstlisting}

\subsubsection{Kernel compilation.}
To build a kernel on the system, some packages are needed before you can successfully build.
You can get these installed with:
\begin{lstlisting}
sudo apt install `\\`libncurses-dev gawk flex bison openssl libssl-dev dkms libelf-dev libudev-dev libpci-dev libiberty-dev autoconf llvm
\end{lstlisting}
Now enter the linux repository and enable the configuration for the kernel:
\begin{lstlisting}
cd ./linux
make menuconfig
\end{lstlisting}
Within the kernel configure menu, make sure to enable the following options:
\begin{lstlisting}[frame=tblr, numbers=none ]
# To enable ByteFS select the following
File systems ---> 
    `\color{red}{<*>}` BYTEFS: ByteFS
# To enable baseline file systems in baseline branch:
File systems ---> 
    `\color{red}{<*>}` The Extended 4 (ext4) filesystem
    `\color{red}{<M>}` F2FS filesystem support
    `\color{red}{<*>}` NOVA: log-structured file system for non-volatile memories
    Miscellaneous filesystems  --->
        `\color{red}{<M>} `Persistent and Protected PM file system support
\end{lstlisting}
We also provide a default configuration file for the kernel under the directory. Note that the provided config may not apply to all setting.

After the configuration, you can build the kernel with the following command:
\begin{lstlisting}
make -j $(threads) 
make modules_install
sudo make install
\end{lstlisting}
Reboot the system to apply the new kernel.

\subsubsection{Reserve DRAM space for emulation.}
Next, to reserve memory region for emulated flash memory, modify the Kernel command line parameters:
\begin{lstlisting}
sudo vim /etc/default/grub # or use any other editor
# add/update grub cmdline: `\\` "GRUB_CMDLINE_LINUX="memmap=nn[KMG]!ss[KMG]"
sudo update-grub
\end{lstlisting}
The cmdline \texttt{memmap=nn[KMG]!ss[KMG]} sets the region of memory to be used, from \texttt{ss} to \texttt{ss+nn}, and \texttt{[KMG]} refers to kilo, mega, giga
(e.g., \texttt{memmap=4G!12G} reserves 4GB of memory between 12th and 16th GB). Configuration is done within GRUB and varies between Linux distributions. In the default setting, set \texttt{nn=32G}.

After rebooting the system, you can check the reserved memory region with:
\begin{lstlisting}
lsblk
\end{lstlisting}
You should see a new device named \texttt{pmem0} with the set size.

\subsubsection{Setting up the workloads}
Run the following commands to install the necessary workloads:
\begin{lstlisting}
sudo ./setworkloads.sh
\end{lstlisting}

\subsection{Evaluation and expected results}
After setting up the experiment environment, you can run specific benchmarks with the following command:
\begin{lstlisting}[language=bash]
    cd ./utils
    # run a given workload using the filesystem specified, and output to designated folder
    ./benchmark_run -w ${workload} -f ${filesystem} -t ${output_folder}
\end{lstlisting} 
The detail options and instructions can be found in the help message of the script with \texttt{-h} option.

Because of different workload settings, we may require rebooting of the machine in between different run configurations. We also offers a script that runs all the experiments and data gathering by executing:
\begin{lstlisting}[language=bash]
    cd ./utils
    # execute benchmark_run.sh on specified remote_target via ssh, and copy output back to local designated folder
    ./run_all.sh -r ${remote_target} -s ${workload_set} -f ${filesystem} -t ${output_folder} 
\end{lstlisting}
The detail options and instructions can be found in the help message of the script with \texttt{-h} option. The experiment result will be in the output folder.


More instructions on how to run the experiments can be found in the README file in GitHub repository.

\subsection{Experiment customization}
\begin{enumerate}[leftmargin=*]
    \item \textbf{Changing workload settings.} Users can customize their own workload by modifying the provided configurations under "./utils/filebench\_workloads" or\\ "./utils/ycsb\_workloads" and evaluate them.
    \item \textbf{Changing emulated SSD capacity.} Users can change the emulated SSD capcity by modifying the kernel command line parameters reserving more DRAM space to perform the experiments. The user should also align the parameter in "./linux/ssd/ftl.h", changing the physical address range (BYTEFS\_PA\_START and BYTEFS\_PA\_END) and also flash parameters (CH\_COUNT, WAY\_COUNT, \\BLOCK\_COUNT, and PG\_COUNT). To apply the changes, re-compiling the kernel.
    \item \textbf{Changing SSD timing model.} The user can change the SSD timing model in "./linux/ssd/timing\_model.h". Changes will be applied after re-compiling the kernel.
    \item \textbf{Changing SSD log size.} The user can change the SSD log size in "./linux/ssd/ftl.h". Changes will be applied after re-compiling the kernel.
\end{enumerate}

More details on how to customize the experiments can be found in the README file in github repository.

\subsection{Methodology}

Submission, reviewing and badging methodology:

\begin{itemize}[leftmargin=*]
  \item \url{https://www.acm.org/publications/policies/artifact-review-badging}
  \item \url{http://cTuning.org/ae/submission-20201122.html}
  \item \url{http://cTuning.org/ae/reviewing-20201122.html}
\end{itemize}



\end{document}